\documentclass[aps,prl,floatfix,english,showpacs,10pt,twocolumn,superscriptaddress]{revtex4-1} %
\pdfoutput=1
\usepackage{amsmath,bm}
\usepackage{mathtools}
\usepackage{commath}
\usepackage{amssymb}
\usepackage{graphicx}
\usepackage{babel}
\usepackage{cases}
\usepackage[colorlinks=true, citecolor=green, anchorcolor=green]{hyperref}
\hypersetup{linkcolor=red}
\usepackage{natbib}
\usepackage{floatrow}
\usepackage{dblfloatfix}
\usepackage{xcolor}
\usepackage{braket}
\usepackage{enumitem}
\usepackage{tabularx}

\setlist[enumerate]{itemsep=0mm}

\makeatletter

\newcommand{\Rmnum}[1]{\expandafter\@slowromancap\romannumeral #1@}

\newcommand{\ext}{\text{ext}}

\newcommand{\de}{\text{de}}
\newcommand{\ff}{\text{fF}}
\newcommand{\ghz}{\text{GHz}}

\makeatother

\begin{document}

\title{A Charge-Noise Insensitive Chiral Photonic Interface for  Waveguide Circuit QED}
\author{Yu-Xiang Zhang}
\email{iyxz@nbi.ku.dk}
\affiliation{Center for Hybrid Quantum Networks (Hy-Q), The Niels Bohr Institute, University of Copenhagen, Blegdamsvej 17, 2100 Copenhagen {\O}, Denmark}
\author{Carles R. i Carceller}
\affiliation{Department of Physics, Technical University of Denmark, Fysikvej 307, 2800 Kgs. Lyngby, Denmark}
\author{Morten Kjaergaard}
\affiliation{Center for Quantum Devices, Niels Bohr Institute, University of Copenhagen, 2100 Copenhagen {\O}, Denmark}
\author{Anders S. S{\o}rensen}
\affiliation{Center for Hybrid Quantum Networks (Hy-Q), The Niels Bohr Institute, University of Copenhagen, Blegdamsvej 17, 2100 Copenhagen {\O}, Denmark}

\begin{abstract}
A chiral photonic interface is a quantum system that has different probabilities for emitting photons to the left and right. 
An on-chip compatible chiral interface is attractive for both fundamental studies of light-matter interactions and applications to quantum information processing.
We propose such a chiral interface based on superconducting circuits, which
has wide bandwidth, rich tunability, and high tolerance to
fabrication variations. The proposed interface consists of a core that uses Cooper-pair-boxes (CPBs) to break time-reversal symmetry, and two superconducting transmons which connect the core to a waveguide in the manner reminiscent of a ``giant atom''. The transmons form a state decoupled from the core, akin to dark states of atomic physics, rendering the whole interface insensitive to the CPB charge noise. The proposed interface can be extended to realize a broadband fully passive on-chip circulator for microwave photons.
\end{abstract}

\date{\today}

\maketitle

The emission of photons from quantum systems is typically non-chiral, meaning that the emitted photons have equal probability to propagate in opposite directions. 
The opposite situation,  where the emitted photons propagate in a single direction,  allows for a wide range of novel phenomena including
cascaded driven-dissipative dynamics~\cite{Gardiner:1993aa,Carmichael:1993aa},
spin dimers~\cite{Ramos:2014aa,Mirza:2016uj},
photonic bound states~\cite{Shen:2007tp,Zheng:2011tf,Mahmoodian:2020aa}, and
solvable models of  waveguide quantum electrodynamics (QED)~\cite{Yudson:1985aa}. 
Chiral interfaces enabling the study of such effects have recently been developed for optical photons~\cite{Lodahl:2017vb},
but optical systems are  prone to losses, limiting the quality of the interfaces. In comparison, superconducting circuits realize almost ideal interfaces to guided microwave photons~\cite{Hoi:2011aa,Blais:2020aa,Roy:2017tg}, potentially allowing for much cleaner demonstrations of these fundamental effects, but so far no chiral interactions have been realized for these systems.

Chiral interfaces to superconducting systems are also of immense technological interest. Scaling superconducting quantum computers to larger sizes puts high demands on the ability to route signals between different components on the chips, e.g., using circulators.
Such circulators necessarily break Lorentz reciprocity and thereby time-reversal symmetry (TRS).
Commercially available circulators often
exploit the Faraday effect~\cite{Caloz:2018aa}. However, such components are off-chip, requiring additional space inside the experimental apparatus.
Most of the existing proposals for on-chip circulators break TRS by tailored active control~\cite{Sliwa:2015aa,Abdo:2014aa,Abdo:2019aa,Kerckhoff:2015aa,Barzanjeh:2017aa,Metelmann:2018aa,Hoi:2011aa,Kamal:2011aa,Estep:2014aa,Roushan:2016aa,Chapman:2019aa},
e.g., via a synthetic magnetic field~\cite{Roushan:2016aa}, or
by dynamically modulating switches and delays~\cite{Chapman:2019aa}.
On-chip circulators that can be operated in a passive form, thereby simplifying its experimental implementations, however remain elusive.
A Josephson junction ring threaded by a constant magnetic flux  breaks TRS~\cite{Koch:2010aa},  but its applicability as a circulator in current superconducting hardware is challenging, because the system must be operated in the Cooper-pair-box (CPB) regime~\cite{Wallraff:2004aa,Koch:2010aa,Muller:2018aa} which is susceptible to charge noise~\cite{Koch:2007aa,Blais:2020aa} and 
limited by bandwidth and fabrication variations~\cite{Carceller:2020ue}. The bandwidth can, however, be increased substantially by exploiting carefully engineered circuits with high impedance \cite{Richman:2020aa}.

\begin{figure}[bt]
\centering
\includegraphics[width=0.95\textwidth]{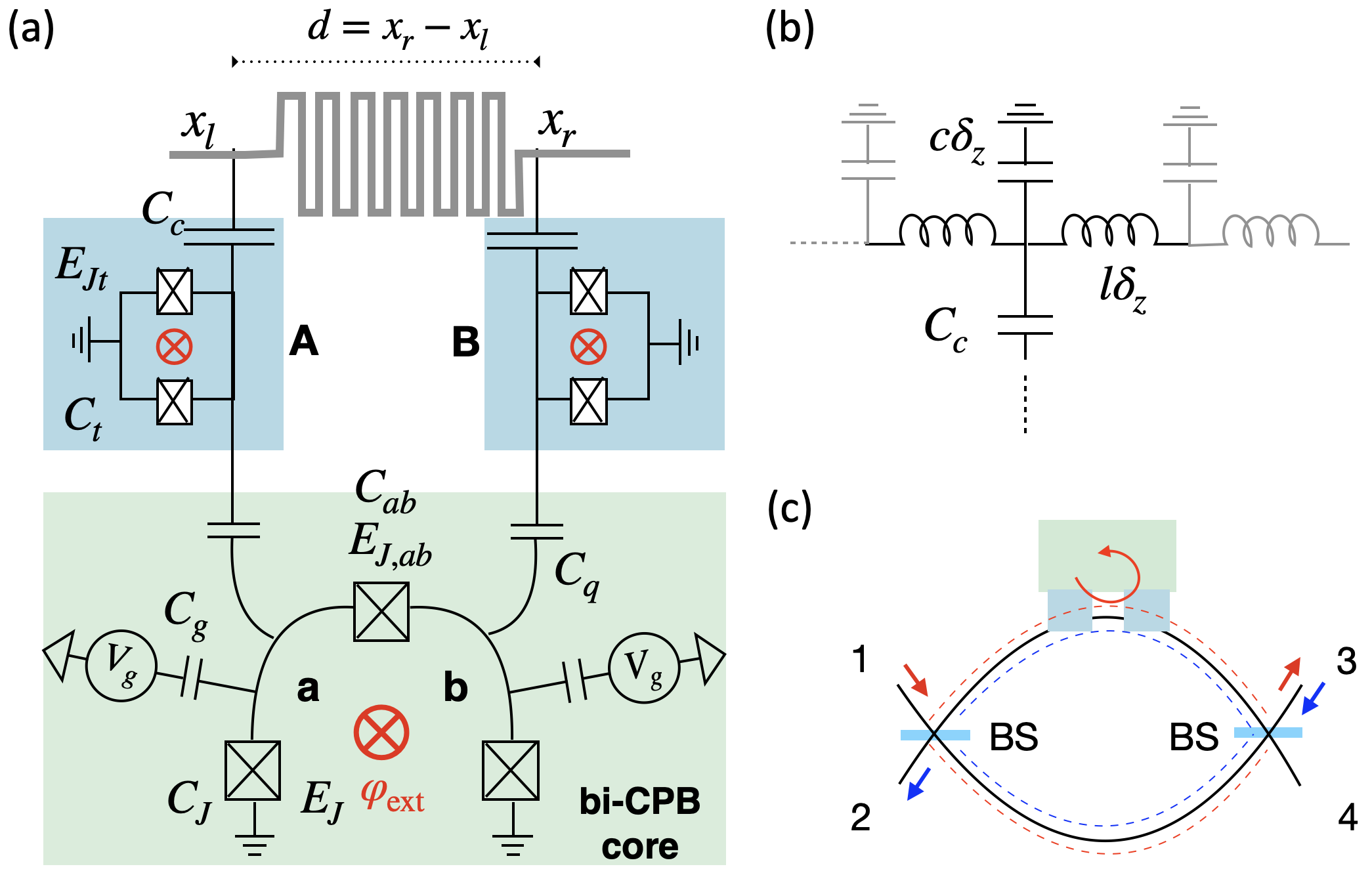}
\caption{Proposal for realizing on-chip compatible charge-noise insensitive chiral photonic interface. (a) Circuit of the chiral interface consisting of two CPBs (green boxes) coupled to a waveguide (top) through two tunable transmons (blue boxes) connecting to the waveguide at positions $x_l$ and $x_r$.  $C_J$ and $E_{J}$ denotes the capacitance and 
Josephson energy of the CPBs, $C_{ab}$ and $E_{J,ab}$ are those of the junction connecting CPB-a and
CPB-b, $\varphi_{\ext}$ is the dimensionless
 external flux, 
$C_g$ is the gate capacitance, $C_q$ couples the CPB and  transmons, $C_t$ and
$E_{J_t}$ are the total capacitance and (tunable) Josephson energy of the transmons
and $C_c$ couples the transmons to the transmission line. 
(b) Circuit model of the
capacitive coupling between a transmon and the transmission line. (c) A circulator based on
chiral interface (right-hand chirality) and Mach-Zehnder interferometer. 
``BS'' denotes  50:50 beam splitters. Input from port-1 goes to port-3 while input from port-3 goes to port-2. }
\label{fig_device}
\end{figure}

In this Letter, we develop an on-chip tunable broadband charge-insensitive chiral system illustrated in Fig.~\ref{fig_device}(a). The system is designed such that the (non-degenerate) excited states decay by only emitting photons in one direction,  left or right.
It has a core composed of two CPBs and an applied external flux (green box),
and two transmons that connect the core to
the waveguide at two separated locations, being reminiscent of a ``giant atom''~\cite{Frisk-Kockum:2014aa,Kockum:2021aa,Kannan:2020aa}. 
The core is essential for breaking TRS, while the transmons amplify the
coupling to the waveguide and hence the bandwidth. 
Furthermore, by operating in a particular dark state configuration~\cite{Arimondo:1976wp}
we show that excitation of the CPBs can be suppressed, such that the interface maintains the excellent coherence properties of transmons~\cite{Place:2021ww,Kjaergaard:2020vo},
while still exploiting the inherent breaking of TRS by the CPBs. 
The interface can directly be extended to realise a circulator for routing signals in large scale superconducting devices.
Finally, the tunability of our chiral interface allows compensating fabrication variations as well as control of the operating frequency within a range of a few GHz.

Recent related proposals 
~\cite{Guimond:2020aa,Gheeraert:2020aa} have used giant atoms formed by transmons and careful adjustments of parameters to realize unidirectional interactions with two degenerate levels, which decay in opposite directions. This degeneracy means that TRS is not broken, and the additional level provides a complication of the dynamics, in particular for non-perfect or partially chiral systems. 
Compared to these works, our chiral interface realizes the paradigmatic chiral model of a two-level system with chiral coupling, has tunable operating frequency, 
and more importantly, can be used for photonic circulators.
During the final stages of this work, a different proposal suggested a chiral interface based on a specially designed waveguide and active driving~\cite{Wang:2021wc}.

\begin{figure}[b]
\centering
\includegraphics[width=\textwidth]{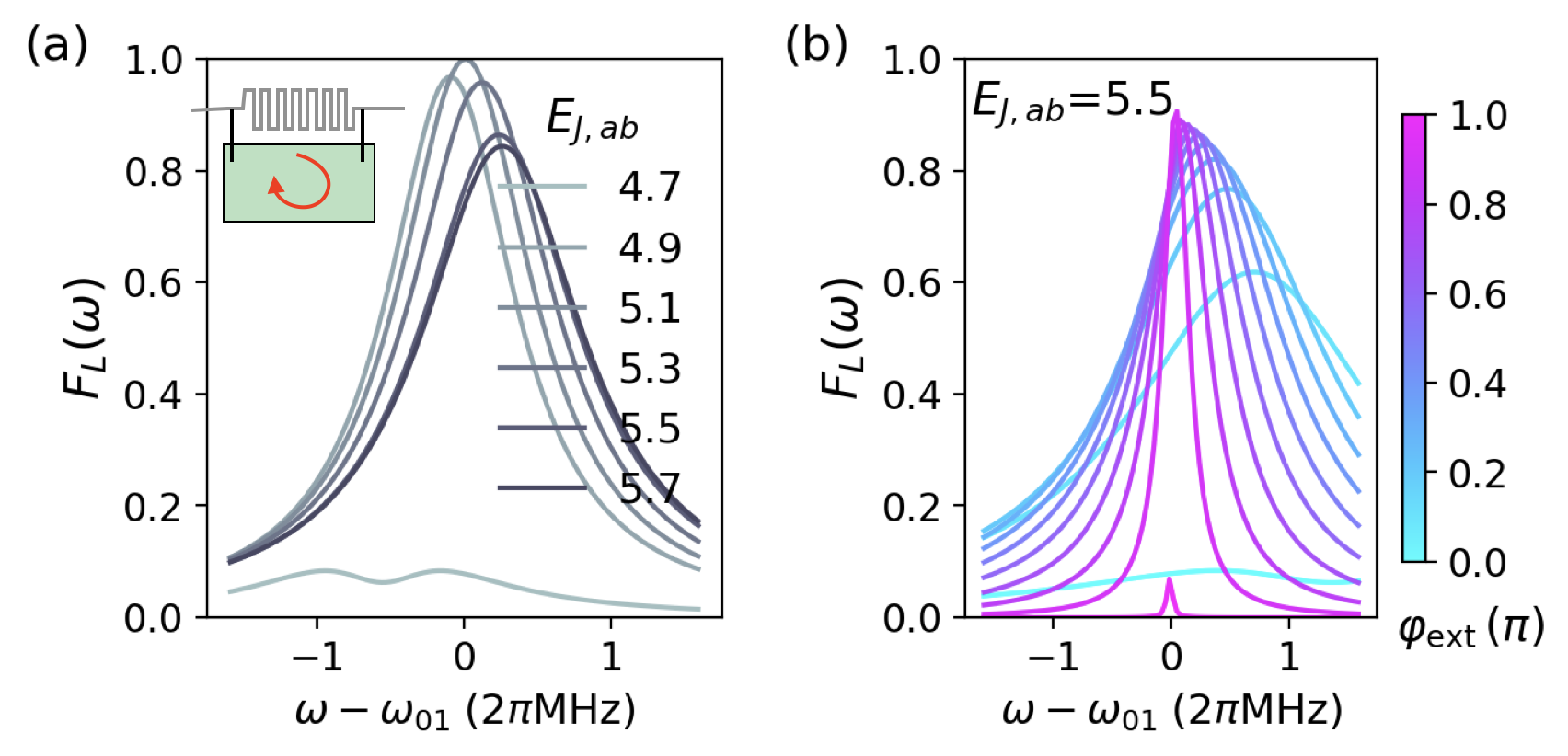}
\caption{Fidelity of circulator $F_L(\omega)$ against the
detuning to the first excited state for direct coupling between the waveguide and the bi-CPB core.
The device parameters are:
$c\delta_z=C_{q}=C_{\Sigma}=C_{ab}=1\ff$, $E_{J}=5.5\, h\ghz$, and
(a) $\varphi_{\ext}=0.5\pi$ and various values of $E_{J,ab}(h\ghz)$;
(b) $E_{J,ab}=5.5\,h\ghz$ and the
external flux $\varphi_{\ext}$ varying from 0 to $\pi$.
}\label{fig_core}
\end{figure}

\paragraph*{Preliminaries.}
We first consider two coupled CPBs, ``a'' and ``b'', which as a whole,
has the ground state $\ket{g_{ab}}$ and a non-degenerate excited state $\ket{e}$.
The transition between them is resonant with waveguide modes 
of wavenumber $\pm k_0$.
We couple CPB-a and CPB-b
to the waveguide at positions $x_l$ and $x_r$, respectively [see Fig.~\ref{fig_device}(a)], resulting in coupling operators of the form 
$\hat{a}_{\pm k_0}(\hat{n}_a+e^{\pm ik_0 d}\hat{n}_b)$, where $\hat{a}_{\pm k_0}$
is the photon annihilation operator,
$d=x_r-x_l$ and $\hat{n}_{a/b}$ is the CPB charge number operator. 
Within the rotating-wave approximation
and perturbation theory, the decay from $\ket{e}$ to left/right going photons depends on the magnitude of 
\begin{equation}
\label{eq-1}
\lambda_{\pm}=\eta_{a}+e^{\pm ik_0d}\eta_b,\quad \eta_{a/b}=\braket{e|\hat{n}_{a/b} |g_{ab}}.
\end{equation}
This formula expresses the interference between excitation decaying through the left and right arms in Fig.~\ref{fig_core}(a), which depends on both the coupling matrix element $\eta_{a/b}$ at $x_{l/r}$ and the phase $e^{\pm ik_0d}$ from waveguide propagation between them.

A chiral interface with a single excited state 
requires TRS  to be broken since left and right propagation are connected by TRS, and a state invariant under TRS cannot distinguish these two possibilities. Breaking of  TRS, however, may induce a phase difference between $\eta_a$ and $\eta_b$ allowing the cancellation of $\lambda_+$ or $\lambda_-$ for a suitable $k_0 d$ if $|\eta_a|=|\eta_b|$.  

Incorporating the chiral interface into one of the arms of a balanced Mach-Zehnder interferometer  enables the construction of a four-port photonic circulator [Fig.~\ref{fig_device}(c)].
Suppose the scattering matrix of the chiral interface is $S_{\epsilon\epsilon'}(\omega_k)$ with
$\epsilon,\epsilon'=\pm 1$ denoting the matrix element from wavenumber $\epsilon' k$ to $\epsilon k$. If the interface has ideal right-hand chirality, there will be
no phase shift for the uncoupled left-moving photons ($S_{--}=1$), whereas a $\pi$ phase shift is attained for the perfectly coupled right-moving photons ($S_{++}=-1$).
This phase difference of $\pi$ implies that the Mach-Zehnder interferometer,
shown in Fig.~\ref{fig_device}(c), can be balanced such that a right-moving photon incident in Port-1 exits through Port-3, whereas a left-moving photon incident in Port-3 exits through Port-2, realizing a circulator. 
We define the fidelity of the circulator as the product of the probabilities for these two inputs to exit through the correct port, which is given by
\begin{equation}\label{fidelity}
F_{R/L}(\omega)=\frac{1}{16}\abs{1\mp S_{++}(\omega)}^2\abs{1\pm S_{--}(\omega)}^2.
\end{equation}
Obviously $F_{R/L}=1$ if $S_{++}= \mp 1$ and $S_{--}= \pm 1$.
In the Supplemental Material~\cite{sp} we show the system Hamiltonian and use the input-output formalism to derive the 
scattering matrix, based on which the fidelity~\eqref{fidelity} can be numerically evaluated for given setups.

\paragraph*{The bi-CPB core.}
TRS can be broken by Josephson junction rings threaded by magnetic flux~\cite{Koch:2010aa}.
Here we use an equivalent ``bi-CPB core''~\cite{sp} depicted in the
green box of Fig.~\ref{fig_device}(a).
It has two CPBs denoted by ``a'' and ``b'', which
are assumed identical for now.
Each CPB has a junction with Josephson energy $E_{J}$ and
capacitance $C_J$. The voltage source determines the offset charge $n_{g,a/b}$
through a capacitor with capacitance $C_g$. We define $C_{\Sigma}=C_J+C_g$.
The two CPBs are connected by a Josephson junction with capacitance $C_{ab}$ and Josephson energy
$E_{J,ab}$. The external flux $\varphi_{\ext}$ is made dimensionless by expressing it in units of $\Phi_0/(2\pi)$ with $\Phi_0$ being the flux quantum. 
As depicted in Fig.~\ref{fig_device}(b), the transmission line waveguide is modeled by an infinite chain of coupled
LC-oscillators with capacitance $c\delta_z$ and inductance $l\delta_z$.  The waveguide
parameters are set by realistic impedance $Z=\sqrt{l/c}=50\Omega$ and
the speed of light $v_g=1/\sqrt{lc}\approx 1.2\times 10^{8} \mathrm{m/s}$.
The elementary length, $\delta_z$, is determined by
the lateral size of the coupling capacitor $C_c$ (or $C_q$ if using only the
bi-CPB core).

At first, we consider coupling the bi-CPB core directly to the waveguide at $x_{l/r}$,
with device parameters given in the
caption of Fig.~\ref{fig_core}. Particularly, we set
$c\delta_z=1\ff$ corresponding to $\delta_z\approx 5\mu m$, which is 
comparable to the size of typical CPBs 
~\cite{Wallraff:2004aa}. To suppress the dephasing caused by charge noise,
the offset charge $n_{g,a/b}$ is fixed at the
``sweet spot'' where $\partial \omega_{01}/\partial n_g=0$~\cite{Vion:2002aa}  and
$\omega_{0n}$ is the energy gap between the $n$th excited 
state and the ground state of the core (the sweet spot $n_{g,a/b}=0.5$ must be 
excluded, because here particle-hole symmetry preserves TRS).
The separation $d = x_r - x_l$ is selected so that $k_0 d=\pi/2$.

In Fig.~\ref{fig_core}(a), we plot $F_L(\omega)$ as a function of detuning from $\omega_{01}$ 
for $\varphi_\ext = 0.5\pi$ and a few values of $E_{J,ab}$. 
The plot shows that the bare core can indeed acts as a chiral emitter, but suffers from several shortcomings.
Firstly, the peak fidelity, $\max_{\omega} F_{R/L}(\omega)$, is sensitive to the exact value of $E_{J,ab}$, leaving the device performance highly sensitive to fabrication variation.
Fabrication variations might be compensated by adapting the tunable flux $\varphi_\ext$. To examine this,
we choose a representative $E_{J,ab}=5.5\,h\ghz$ corresponding to a 10\% deviation from the ideal value 
and plot $F_L(\omega)$ as a function of $\varphi_{\ext}$ in Fig.~\ref{fig_core}(b). It shows that
this tuning fails to make the maximal fidelity above 0.9.
Furthermore, tuning $\varphi_\ext$ changes $\omega_{01}$, rendering this configuration inadequate for tasks where photon frequencies are fixed. 
Figure~\ref{fig_core} also shows that
the bandwidth  is less than
1$\mathrm{MHz}$, which severely limits the functionality.

\begin{figure}[b]
\centering
\includegraphics[width=0.85\textwidth]{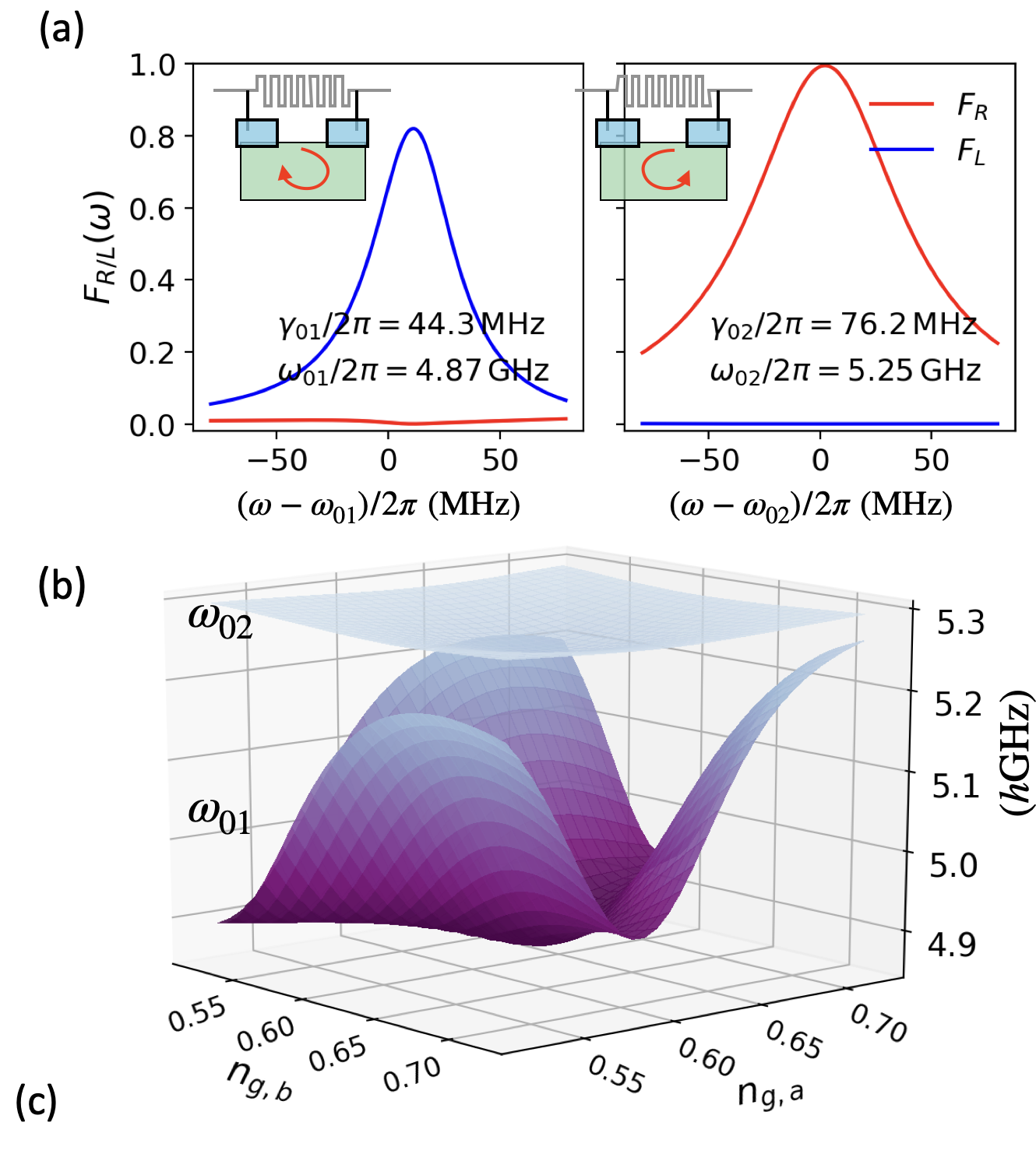}
\caption{Energy levels and performance of the full system. (a) Fidelity of circulator as a function of the detuning from $\omega_{01}$
and $\omega_{02}$ with device parameters in (c). (b)Values (in unit of $h\mathrm{GHz}$) of $\omega_{01}$ and $\omega_{02}$ 
of the full device as a function of $n_{g,a/b}$, using parameters listed in (c), where units for capacitances and Josephson energies are $\ff$ and $h\ghz$, respectively.
}\label{fig_levels}
\begin{tabularx}{0.8\textwidth}{c c| c c c c c c | c c c c}
\hline
\hline
$c\delta_z$ &$k_0d$ &$C_q$ &$C_{\Sigma}$ &$C_{ab}$  &$E_{J}$ &$E_{J,ab}$ &$\varphi_{\ext}$
&$C_c$ &$C_t$ &$E_{Jt}$ \\
\hline
50 &0.5$\pi$ &1 &1 &1 &5.5 &5 &0.5$\pi$  &100  &30 &13\\
\hline
\hline
\end{tabularx}
\end{figure}

\paragraph*{The transmon-coupled bi-CPB chiral photonic interface.}
The bandwidth is limited by the weak interaction between the small CPBs and the waveguide excitations delocalized on wavelength scale. 
To overcome this, we propose to insert transmons,
which have lateral size in between the CPBs and the photonic wavelength, as efficient mediators.
Compared with
similar proposals using microwave resonators~\cite{Koch:2007aa,Richman:2020aa},
transmons allow for stronger spatial confinement of excitations, and thereby larger couplings to the CPBs, while simultaneously maintaining sizable coupling to waveguides. A similar advantage could be achieved with high impedance elements \cite{Richman:2020aa}, but the transmons additionally allow wide tunability as discussed below.
In Fig.~\ref{fig_levels}(a) we plot $F_{R/L}(\omega)$ as a function of
the detuning to the first two excited states ($\omega_{01}$ and $\omega_{02}$)
for the CPB-transmon-coupled system. The device parameters used
in the calculation are listed in
Fig.~\ref{fig_levels}(c), which are chosen to be representative for typical experimental situations and not fully  optimized. 
Particularly,
we have taken $c\delta_z=50\ff$ i.e., $\delta_z\approx 250\mathrm{\mu m}$, corresponding to the lateral size of the transmon reported in Ref.~\cite{Mirhosseini:2019aa}.
Figure~\ref{fig_levels}(a) shows that the bandwidth is significantly improved.
However, while a left-hand chirality ($F_{L}\gg F_{R}$) is realized for $\omega\approx\omega_{01}$ (and $\omega\approx\omega_{03}$) -- similar to the bare bi-CPB core -- we observe
the opposite chirality for $\omega\approx \omega_{02}$ ($F_{R}\gg F_{L}$).
This feature indicates the distinctness of the 2nd excited state.
Below we use an approximate 
model to highlight 
multiple benefits brought by this state.

We assume that the two transmons, indexed by ``A'' and ``B'' in 
Fig.~\ref{fig_device}(a), are identical in parameters, and consider
a restricted Hilbert space spanned
by the ground state and the singly-excited states of either transmon-A,B, or the
bi-CPB core, which will be denoted by $\ket{1_A}$, $\ket{1_B}$,
and $\ket{e}$, respectively. The device Hamiltonian reads
$H=H_0+H_{\text{int}}$, where
$H_0=e_1\ket{e}\bra{e}+\omega_t(\ket{1_A}\bra{1_A}+\ket{1_B}\bra{1_B})$
with $e_1$ and $\omega_t$ the energies of the 1st excited states of the core
and the transmons, respectively. The interaction Hamiltonian is
\begin{equation}\label{H-model}
H_{\text{int}}=g\hat{n}_A(\hat{n}_a -n_{g,a})+
g\hat{n}_B (\hat{n}_b -n_{g,b}),
\end{equation}
where $g$ is the coupling strength.
Using rotating-wave approximation upon $H_\text{int}$, and restricting it
to the reduced Hilbert space, we rewrite $H$ as
\begin{equation}
    H=H_0+\tilde{g}\big(\ket{\psi_c}\bra{e}
    +\ket{e}\bra{\psi_c}\big),
\end{equation}
where $\tilde{g}$ is modified from $g$ and
\begin{equation}\label{psic}
\ket{\psi_{\text{c}}} \propto\eta_a^{*}\ket{1_A}+\eta_b^{*}\ket{1_B},
\end{equation}
with $\eta_{a(b)}=\braket{e|\hat{n}_{a(b)}|g_{ab}}$ as in Eq.~\eqref{eq-1}.
The coupling between $\ket{\psi_c}$ and $\ket{e}$ yields
the 1st excited state $\ket{\Psi_{-}}$ with energy $\omega_{01}=e_1+\Delta_-$
and the 3rd excited state $\ket{\Psi_{+}}$ with energy $\omega_{03}=e_1+\Delta_+$,
\begin{equation}
\ket{\Psi_{\pm}} \propto \underbrace{\Delta_{\pm}\ket{\psi_{\text{c}}}}_{\text{transmons}}+\underbrace{i\tilde{g}\ket{e}}_{\text{bi-CPB}}
\end{equation}
where $\Delta_{\pm}=-\delta/2\pm\sqrt{\delta^2/4+\tilde{g}^2}$ with $\delta=e_1-\omega_t$.
The 2nd excited state is a so-called dark state~\cite{Arimondo:1976wp}, which has no excitation on the core,
\begin{equation}\label{pside}
\ket{\Psi_{\text{de}}}\propto \eta_{b}\ket{1_A}-\eta_a\ket{1_B}.
\end{equation}
Despite being decoupled, $\ket{\Psi_{\de}}$ still attains chiral property from the core. The constants  $\eta_{a(b)}$, however,  have
different orders and signs in $\ket{\Psi_{\text{de}}}$ and $\ket{\psi_c}$, which lead to opposite chirality.

A more important observation is that the
 state $\ket{\Psi_{\text{de}}}$ 
does not contain the bi-CPB state $\ket{e}$. Thus, it is not affected by the CPB charge noise and keeps the charge noise insensitivity of the transmon qubits~\cite{Koch:2007aa}.
To illustrate this, Fig.~\ref{fig_levels}(b)  shows $\omega_{01}$ and $\omega_{02}$ as a function of the offset charges $n_{g,a/b}$. The results are numerically evaluated from the complete Hamiltonian (see~\cite{sp} for details) using device parameters listed in~Fig.~\ref{fig_levels}(c). 
Figure~\ref{fig_levels}(b) demonstrates that $\omega_{02}$ is significantly more 
stable than $\omega_{01}$.
The fluctuation of $\omega_{02}$ does not exceed its linewidth $\gamma_{02}/2\pi\approx 76\,\mathrm{MHz}$ despite the large variations in $n_{g,a/b}$.
Thus, the offset charges $n_{g,a(b)}$ need not be
at the sweet spots, and can be used as control knobs to provide  tunability.

\begin{figure}[!t]
\centering
\includegraphics[width=\textwidth]{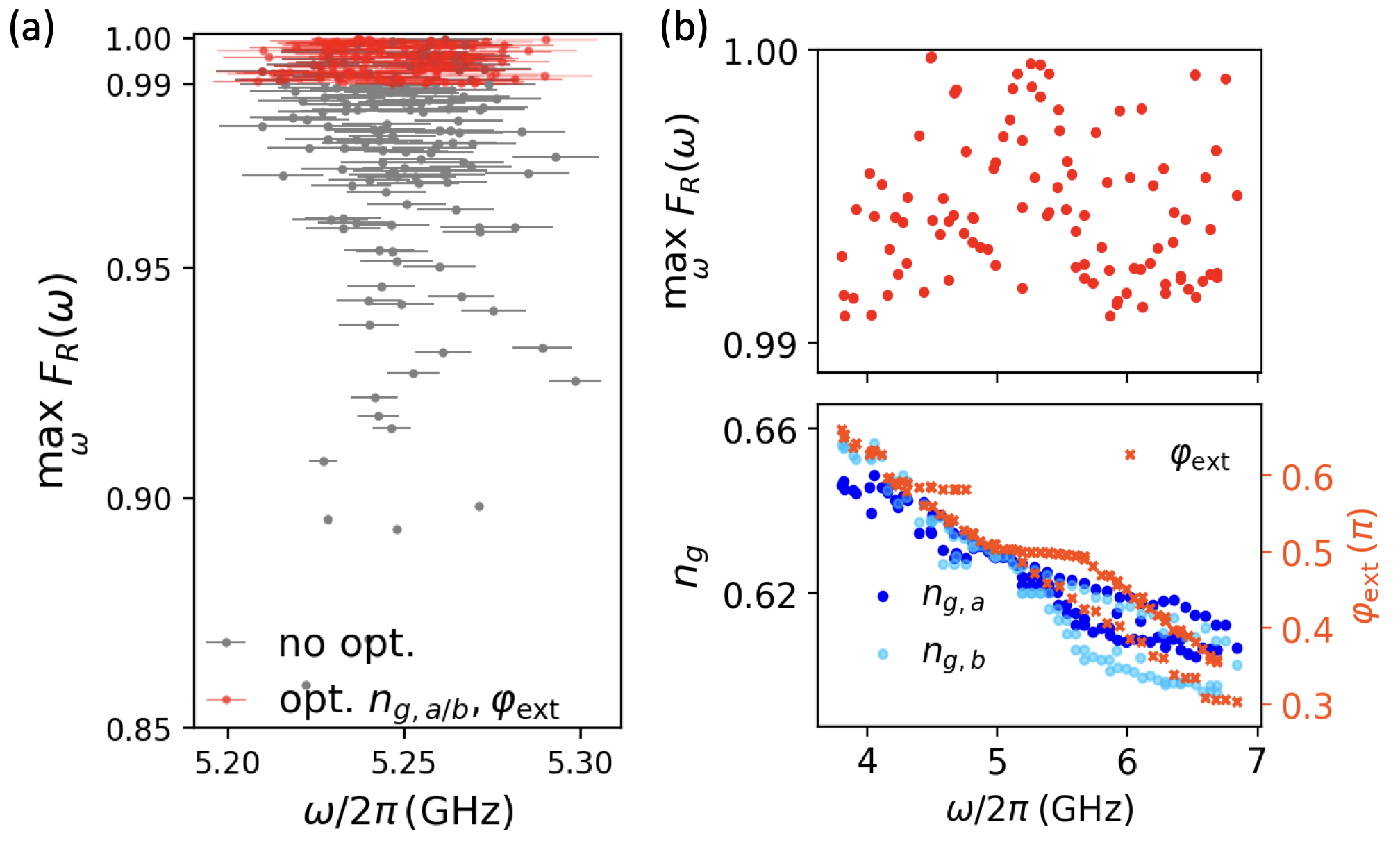}
\caption{Tunability of the chiral interface. (a) 150 realizations of the chiral interfaces with device parameters randomly sampled to represent  fabrication variations (see text for details). The marker shows peak fidelity for a given realization, error bar indicates the $F_{R}>0.9$ bandwidth.
The grey points show results obtained when $n_{g,a/b}$ and $\varphi_\ext$ are optimized for the ideal parameters listed in Fig.~\ref{fig_levels}(c).
Red points show fidelities obtained when $n_{g,a/b}$ and $\varphi_\ext$ are optimized for each realization. (b) upper panel: Peak fidelities as a function of photon frequency, for the chiral interface with
parameters given in Fig.~\ref{fig_levels}(c) and 100 samples of transmon Josephson energy $E_{Jt}$ between $6\sim 22\, h\ghz$.
For each red dot thereof, the vertically aligned markers in
the lower panel shows the adaptively optimized $n_{g,a/b}$ and $\varphi_\ext$.}
\label{fig_mc}
\end{figure}

\paragraph*{Tunability of the chiral interface.}
First, we demonstrate that by tuning $n_{g,a/b}$ and $\varphi_\ext$
we can address issues of fabrication variation in the device parameters. To simulate realistic fabrication variations,
we sample values of capacitance and CPB
Josephson energy uniformly in intervals bounded by $\pm 1\%$ and $\pm 10\%$,
respectively, from the intended values listed in Fig.~\ref{fig_levels}(c).  The transmon Josephson energy $E_{Jt}$ 
is fixed at the intended values
because they can be tuned \emph{in situ} via flux-control.
We collect 150 such ``realistic samples'', which includes
cases with asymmetries between CPB-a and CPB-b, and between transmon-A and transmon-B.
In Fig.~\ref{fig_mc}(a), we fix $n_{g,a/b}$ and $\varphi_\ext$ at the optimal value for the ideal parameters, and show by grey points their individual maximal fidelity (vertical coordinate), frequency where the maximal fidelity is obtained (horizontal coordinate), and the $F_{R/L}>0.9$ bandwidth (horizontal error bar).
While the performance is generally good, certain instances with poor chiral fidelity are observed.
Next, for each sample that has a maximal fidelity less than 0.99, we search for values of $n_{g,a/b}$ and $\varphi_\ext$ that improve the performance.
The red points in Fig.~\ref{fig_mc}(a) show the result of such optimization. We are always able to tune the maximal fidelity above $0.99$.
Additional details about how the maximal fidelity and bandwidth depend on $n_{g,a/b}$ and $\varphi_\ext$ are given in the Supplemental Material~\cite{sp}.

Additionally, flux-tunable transmons~\cite{Koch:2007aa} enables tuning of the operating frequency of our device.
We consider the parameters listed in Fig.~\ref{fig_levels}(c) (including the separation $d$). For 100 values of $E_{Jt}$ between $6\sim 22\, h\ghz$, we optimize the peak fidelity to be higher than 0.99 by tuning $n_{g,a/b}$ and $\varphi_\ext$. The results are presented in Fig.~\ref{fig_mc}(b). 
It shows that for photons in a band of $3\ \ghz$ width, the same device can always be tuned to provide an excellent chiral interaction.
This broad adaptability is
rooted in the breaking of TRS. 
When we tune $E_{Jt}$, $\omega_{02}$ changes and so does the
phase $k_0d$ with $k_0=\omega_{02}/v_g$. Adjusting the flux $\phi$, however, allows us to achieve complete destructive interference in one of the directions. 
Our chiral interface thus works in principle as long as $k_0 d\neq 0$ ($\mathrm{mod}\,\pi$). 

\paragraph*{Discussion and conclusion.} 
The bandwidth for $F_{R/L}(\omega)>0.9$
is about $25\,\mathrm{MHz}$ in Fig.~\ref{fig_levels}(a). This is limited by the transmon decay rate, which is typically
on the order of $100\,\mathrm{MHz}$. The bandwidth can thus be directly enhanced by designing transmons with larger decay rates. We note however that as the  decay rate approaches roughly a tenth of the transition frequency ($0.1\omega_0/2\pi\sim 500\mathrm{MHz}$) the system reaches the ultra-strong coupling regime of waveguide QED~\cite{Forn-Diaz:2016aa} requiring a more advanced theoretical description~\cite{Sanchez-Burillo:2014aa,Shi:2018aa}.

Our interface relies on the 2nd excited state. This raises the concern that  decay to the 1st excited state may jeopardize the performance.  This transition is negligible for two reasons. First, the transition
matrix element is tiny (vanishing in our simplified model
$\braket{\Psi_-|\hat{n}_{A/B} |\Psi_{\de}}=0$). Second, 
the waveguide mode density at $\omega_{12}/2\pi=(\omega_{02}-\omega_{01})/2\pi\sim 100\mathrm{MHz}$
is low. As a result, the decay rate to the 1st excited state is numerically found to be  $\gamma_{12}/2\pi\sim 10\,\mathrm{kHz}$, 
which is much smaller than $\gamma_{02}/2\pi\sim 100\,\mathrm{MHz}$. 
Similarly, the intrinsic loss of the components involved can be made well below $0.1\mathrm{MHz}$~\cite{Barends:2013tl,Place:2021ww}. This is much less than the
decay rate to the waveguide and hence negligible~\cite{sp}.

In conclusion, we have shown that by introducing two transmons, a passive charge-noise insensitive chiral photonic interface can be build, based on a core consisting of CPBs and an external magnetic flux that breaks TRS.
The interface has rich tunability, making it tolerant to experimentally relevant fabrication variations.
The bandwidth of $F_{R/L}>0.9$ is a few tens of $\mathrm{MHz}$ for realistic transmon parameters with decay rates typically around $100\mathrm{MHz}$. With tunable transmons (SQUIDs), our design can be tuned in a band as wide as a few $\ghz$. The lateral dimension of the entire circulator can be a few millimeters~\cite{sp}.
These properties makes our chiral interface highly attractive for e.g., on chip routing of microwave signals~\cite{Sliwa:2015aa,Abdo:2014aa,Abdo:2019aa,Kerckhoff:2015aa,Barzanjeh:2017aa,Metelmann:2018aa,Hoi:2011aa,Kamal:2011aa,Estep:2014aa,Roushan:2016aa,Chapman:2019aa,Guimond:2020aa,Gheeraert:2020aa} as well as fundamental studies of chiral waveguide QED~\cite{Ramos:2014aa,Mirza:2016uj,Mahmoodian:2018aa,Mahmoodian:2020aa}.

\begin{acknowledgments}
A. S. S{\o}rensen thanks Peter Lodahl, Per Delsing and Göran Johansson for useful discussions. 
M. Kjaergaard gratefully acknowledges helpful and insightful discussions with Bharath Kannan. Y.-X. Zhang and A. S. S{\o}rensen thank Jacob M. Taylor and Brittany Richman for insightful discussions.
Y.-X. Zhang, M. Kjaergaard \& A. S. S{\o}rensen acknowledge financial support from the Danish National Research Foundation.
M. Kjaergaard was also supported by Villum Fonden (grant 37467) through a Villum Young Investigator grant. 
\end{acknowledgments}

\bibliography{circuitQED}

\end{document}


\title{Supplemental Material for ``A Charge-Noise Insensitive Chiral Photonic Interface for  Waveguide Circuit QED''}

\author{Yu-Xiang Zhang}
\affiliation{Center for Hybrid Quantum Networks (Hy-Q), The Niels Bohr Institute, University of Copenhagen, Blegdamsvej 17, 2100 Copenhagen {\O}, Denmark}
\author{Carles R. i Carceller}
\affiliation{Department of Physics, Technical University of Denmark, Fysikvej 307, 2800 Kgs. Lyngby, Denmark}
\author{Morten Kjaergaard}
\affiliation{Center for Quantum Devices, Niels Bohr Institute, University of Copenhagen, 2100 Copenhagen {\O}, Denmark}
\author{Anders S. S{\o}rensen}
\affiliation{Center for Hybrid Quantum Networks (Hy-Q), The Niels Bohr Institute, University of Copenhagen, Blegdamsvej 17, 2100 Copenhagen {\O}, Denmark}

\date{\today}

\begin{abstract}
This Supplemental Material is organized as follows: 
In Sec.~\ref{theory}, we derive the Hamiltonians of the chiral interface and
the coupling to the transmission line waveguide, and present the Hilbert space truncation used 
in all the numerical calculations.
In Sec.~\ref{equivalence-3CPB}, we discuss the equivalence between our bi-CPB core
and the Josephson junction ring proposed in Ref.~\cite{Koch:2010aa}.
In Sec.~\ref{scattering matrix}, we derive the single-photon scattering matrix
of the chiral interface using the standard input-output formalism. 
In Sec.~\ref{dev-3}, we show data parallel to that presented in the main text, but for a different family of device parameters.
In Sec.~\ref{more_data}, we calculate the maximal fidelity
and bandwidth of $F_{R/L}>0.9$ for different values of the CPB offset charges $n_{g,a(b)}$ and constant external 
flux $\varphi_\ext$, in order to demonstrate that tuning  $n_{g,a(b)}$ and $\varphi_\ext$
is able to circumvent fabrication variation. In Sec.~\ref{size} we
estimate the geometric size of the whole device, and the influence of
inherent decay of the device to the fidelity of chirality. 
\end{abstract}

\maketitle

\section{Theoretical Formalism for the Chiral Interface}
\label{theory}

In this section,  we introduce the Hamiltonian of the chiral interface,
including the artificial atoms and the transmission line waveguide. 
This section is organized as follows: In Sec.~\ref{quantize_wg} we 
present a discrete model for the transmission line, which can be conveniently
integrated into the waveguide-``giant atom'' coupling.
In Sec.~\ref{HCI} we elaborate on the Hamiltonian of the
artificial giant atom. In Sec.~\ref{remark_num}
we  present the way we numerically diagonalize the Hamiltonian
of the artificial atom.

\subsection{Quantization of the Transmission Line Waveguide}\label{quantize_wg}
We assume that the transmission line is infinite.
In the literature, the one-dimensional continuum of waveguide modes
are usually described through a continuous flux field $\phi(x)$. The Lagrangian of the waveguide reads
\begin{equation}
L_{tr}=\int_{-\infty}^{\infty} dx\; \bigg[\frac{1}{2}c\dot{\phi}^2-\frac{1}{2l}(\partial_x\phi)^2 \bigg].
\end{equation}
where $c$ and $l$ are the  capacitance and inductance per unit length.

\begin{figure}[b]
\centering
\includegraphics[width=0.9\textwidth]{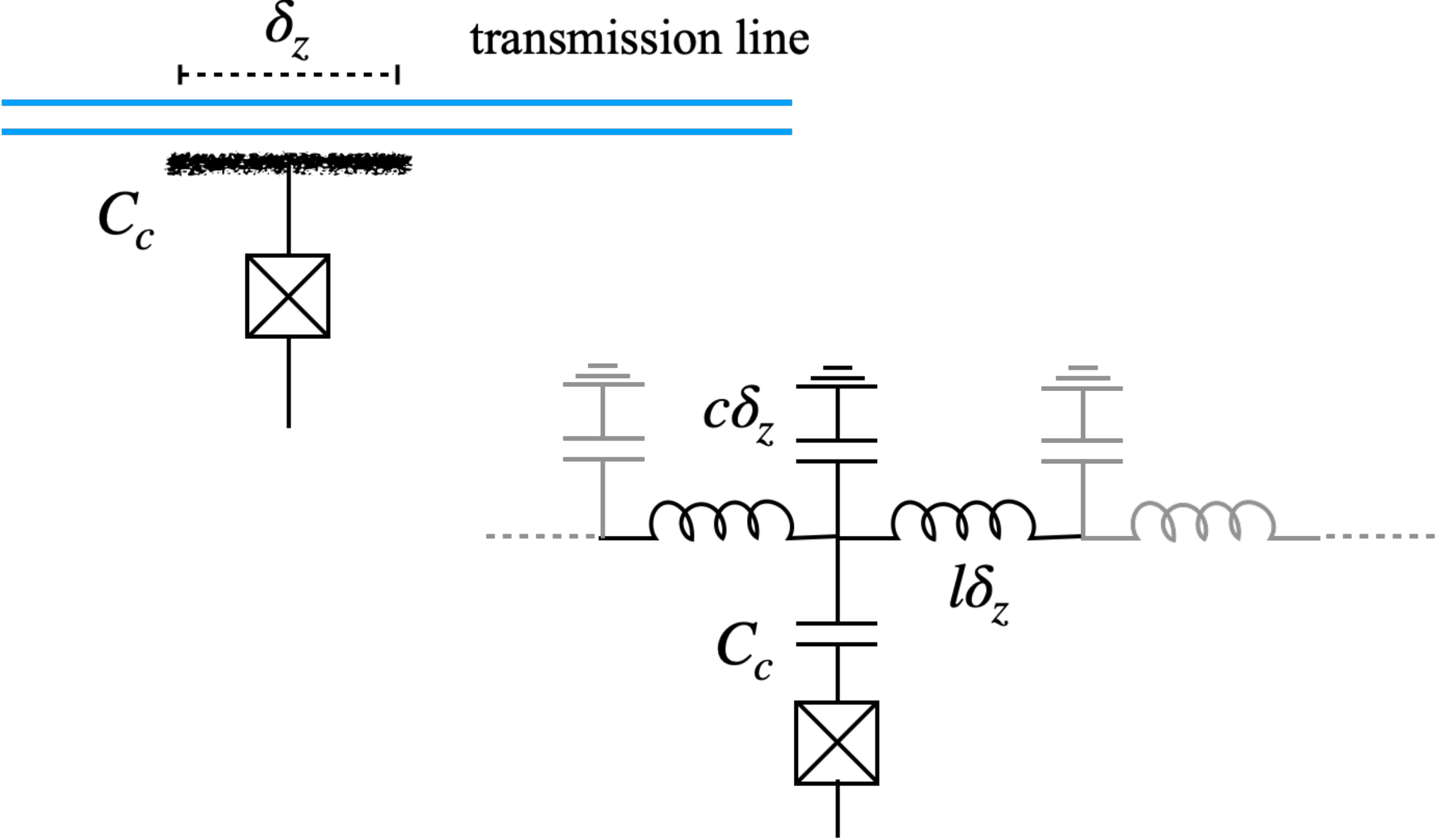}
\caption{Capacitive coupling to the transmission line. Upper panel:continuous model; 
Lower panel: discrete model.}
\label{fig_tline}
\end{figure}

For our purpose it is convenient to describe the waveguide as a discrete series of 
coupled LC-oscillators. In this model
each oscillator has capacitance $c\delta_z$
and inductance $l\delta_z$ with $\delta_z$  an infinitesimal unit length. 
The coupling capacitor has finite physical length,
as illustrated in Fig.~\ref{fig_tline}. Since the coupling cannot 
distinguish modes with wavelengths shorter than this length, it is convenient to
define $\delta_z$ to be this length, so that
the variable  relevant to the coupling is
\begin{equation}
\bar{\phi}(x_{l/r})=\frac{1}{\delta_z} \int_{x_{l/r}-\delta_z/2}^{x_{l/r}+\delta_z/2}dx \; \phi(x).
\end{equation}
We then obtain a discrete model:
\begin{equation}
L_{tr}=\sum_{n=-\infty}^{+\infty} \bigg[
\frac{1}{2}c\delta_z \dot{\phi}_{n}^2-\frac{1}{2l\delta_z }(\phi_{n+1}-\phi_{n})^2
\bigg]
\end{equation}
The corresponding Hamiltonian is given by
\begin{equation}\label{transmission-H}
H_{tr}=\sum_{n=-\infty}^{\infty} \bigg[
\frac{1}{2c\delta_z} q_n^2+\frac{1}{2 l\delta_z} (\phi_{n+1}-\phi_{n})^2\bigg].
\end{equation}

To diagonalize the Hamiltonian, we move to momentum representation:
\begin{equation}
\begin{aligned}
\phi_m & =\int_{-\pi}^{\pi} \frac{dk}{2\pi} \phi_k e^{im k } \\
q_m & =\int_{-\pi}^{\pi} \frac{dk}{2\pi} q_k e^{-im k},
\end{aligned}
\end{equation}
Since $[\phi_m, q_n]=i\hbar\delta_{mn}$, we have $[\phi_k, q_{k'}]=i\hbar 2\pi\delta(k-k')$. 
Note that $k$ is here defined to be dimensionless.
Substituting the above equations into Eq.~\eqref{transmission-H} and using the equality
$\sum_{n=-\infty}^{\infty} e^{-i k n}=2\pi\delta(k)$, we get
\begin{equation}
H_{tr}=\frac{1}{2}\int_{-\pi}^{\pi}\frac{dk}{2\pi}\bigg( \frac{q_{k}q_{-k}}{c\delta_z} +
4|\sin\frac{k}{2}|^2 \frac{\phi_k \phi_{-k}}{l\delta_z}\bigg).
\end{equation}
Next, we introduce bosonic ladder operators $b_k$ and $b_k^{\dagger}$ satisfying 
$[b_k, b_{k'}^{\dagger}]=2\pi\delta(k-k')$, and
\begin{subequations}
\begin{equation}
\phi_k=\sqrt{\frac{\hbar Z_0}{4\abs{\sin(k/2)}}} (b_k + b_{-k}^{\dagger}),
\end{equation}
\begin{equation}
q_k=i\sqrt{\frac{\hbar\abs{\sin(k/2)}}{Z_0}} (b_k^{\dagger} -b_{-k}),
\end{equation}
\end{subequations}
where $Z_0=\sqrt{l/c}$ is the impedance. Then the Hamiltonian is diagonalized to 
$H_{tr}=\int_{-\pi}^{\pi}\frac{dk}{2\pi}\hbar\omega_k b_k^{\dagger} b_k$ with
\begin{equation}
\omega_k=\frac{2}{\sqrt{lc}\delta_z}\abs{\sin(k/2)}.
\end{equation}

The phase and charge are expressed as
\begin{subequations}
\begin{equation}
\phi_m=\frac{1}{2}\int_{-\pi}^{\pi}\frac{dk}{2\pi} \frac{\sqrt{\hbar Z_0}}{\sqrt{|\sin (k/2)|}}\;
(e^{ikx_m} b_k + e^{-ikx_m} b_k^{\dagger}),
\end{equation}
\begin{equation}
q_m=i\int_{-\pi}^{\pi}\frac{dk}{2\pi} \sqrt{\frac{\hbar}{Z_0}|\sin\frac{k}{2}|} \;
( e^{-ikx_m} b_k^{\dagger}- e^{ikx_m} b_k).
\end{equation}
\end{subequations}
The two unit cells connected to the ``giant atom'' will be denoted by ``l'' and
``r''. Their charge number operators $\hat{n}_{l/r}$ are obtained
from the charge operators $\hat{q}_{a/b}$ via the relation 
$\hat{q}_{l/r}=2e\hat{n}_{l/r}$.

\subsection{Hamiltonian of the Chiral Interface}\label{HCI}

At first, we consider a cluster composed by the two transmons, the bi-CPB core, and
two sections of the transmission line, $x_l$ and $x_r$. The Hamiltonian for these parts reads
\begin{equation}
\begin{aligned}
H_{0}=&2e^2\bm{\hat{n}}\bm{C}^{-1}\bm{\hat{n}}-\sum_{i=a,b}E_{J,i}\cos(\hat{\varphi}_i)
-\sum_{j=A,B}E_{Jt,j}\cos(\hat{\varphi}_j) \\
&-E_{J,ab}\cos(\hat{\varphi}_a-\hat{\varphi}_b-\varphi_{\ext})
+\frac{1}{2l\delta_z}(\hat{\phi}_{l}^2+\hat{\phi}_r^2),
\end{aligned}
\end{equation}
where $\bm{\hat{n}}=(\hat{n}_l, \hat{n}_A, \hat{n}_a-n_{g,a}, \hat{n}_r,\hat{n}_B,\hat{n}_b-n_{g,b})^\top$.
The dimensionless flux used in the Josephson potentials 
is defined in the form of $\hat{\varphi}=\hat{\phi}\times 2\pi/\Phi_0$, and the capacitance matrix is given by
\begin{equation}
\bm{C}=\begin{pmatrix}
c\delta_z & -C_{c,A} &0 &0 &0 &0 \\
-C_{c,A} & C_{\Sigma,A}&-C_{q,a} &0 &0 &0\\
0 &-C_{q,a} & C_{\Sigma,a} &0 &0 &-C_{ab} \\
0 &0 &0 &c\delta_z &-C_{c,B} &0\\
0 &0 &0 &-C_{c,B} &C_{\Sigma,B} &-C_{q,b} \\
0 &0 &-C_{ab} &0 &-C_{q,b} &C_{\Sigma,b}
\end{pmatrix}
\end{equation}
where $C_{\Sigma,A/B}=C_{c,A/B}+C_{t,A/B}+C_{q,a/b}$ is the
total capacitance of transmon-A/B, and $C_{\Sigma,a/b}=C_{\Sigma,a/b} +C_{q,a/b}+C_{ab}$ is the
total capacitance of CPB-a/b. Note that offset charges of the transmons are not included. This is justified 
because for the low energy excitations, the influence of offset charges is exponentially suppressed if $E_{Jt}$ is
much larger than the transmon charging energy~\cite{Koch:2007aa}.

 $H_{0}$ can be decomposed into the ``system'' that consists of
transmon-A/B and the bi-CPB core:
\begin{equation}\label{hsys}
\begin{aligned}
H_{sys}=&2e^2\sum_{i,i'=a,b}\bm{C}^{-1}_{i,i'}(\hat{n}_i-n_{g,i})(\hat{n}_{i'}-n_{g,i'})\\
&-\sum_{i=a,b} E_{J,i}\cos(\hat{\varphi}_i)-E_{J,ab}\cos(\hat{\varphi}_a-\hat{\varphi}_b-\varphi_\ext)\\
&+2e^2\sum_{j,j'=A,B}\bm{C}^{-1}_{j,j'}\hat{n}_j\hat{n}_{j'}-\sum_{j=A,B}E_{Jt,j}\cos(\hat{\varphi}_j) \\
&+4e^2\sum_{i=a,b}\sum_{j=A,B} \bm{C}^{-1}_{ij}\hat{n}_j(\hat{n}_i -n_{g,i}),
\end{aligned}
\end{equation}
where terms in the last line covers Eq.~(3) of the main text.
The interactions to
the two unit cells ``l'' and ``r'' of the transmission line:
\begin{equation}\label{hint}
\begin{aligned}
H_{int}=&4e^2\sum_{\xi=l,r}\sum_{i=a,b} \bm{C}^{-1}_{\xi,i}\hat{n}_{\xi}(\hat{n}_i-n_{g,i})\\
&+4e^2\sum_{\xi=l,r}\sum_{j=A,B} \bm{C}^{-1}_{\xi,j}\hat{n}_{\xi}\hat{n}_j.
\end{aligned}
\end{equation}

We neglect the corrections to the 
charging energy from the two
transmission line unit cells, and also neglect their direct interaction (expressed by terms with coefficient $C^{-1}_{l,r}$), which is bridged by the coupled core and transmon system. This means that the transmission line Hamiltonian is $H_{tr}$~\eqref{transmission-H} without  corrections other than from the coupling to the excitations of the systems. That is,
the total Hamiltonian is written as
\begin{equation}
H=H_{sys}+H_{tr}+H_{int}.
\end{equation}

\subsection{Hilbert Space Truncation Used in the Numerical Calculations}\label{remark_num}

\begin{table}[t]
\caption{Energy levels of the chiral interface (in units of $h\ghz$) calculated with different truncations of the Hilbert space. The column 
$\mathcal{H}_t /\mathcal{H}_{\cpb}$ lists the dimensions of the Hilbert spaces of the transmons and the bi-CPB core.
}\label{tab_trunc}  
\begin{ruledtabular}
\begin{tabular}{ c | c c c c c}
\multicolumn{6}{c }{ Dev-1: main text Fig.~3(c)} \\
\hline
$\mathbf{\mathcal{H}_t/\mathcal{H}_{\cpb} }$ &
$\omega_{01}$ & $\omega_{02}$ & $\omega_{03}$ & $\omega_{04}$ & $\omega_{05}$ \\ 
$3^2/10^2$ & 5.3321 & 5.6583 & 6.1254 &9.4978 &10.8150 \\
$4^2/8^2$ & 5.3292 & 5.6640 & 6.1222 &9.4998 &10.7921 \\
$4^2/10^2$ & 5.3292 & 5.6640 & 6.1222 &9.4998 &10.7921 \\
$5^2/12^2$ & 5.3294 & 5.6636 & 6.1224 &9.4997 &10.7941  \\  
\hline
\hline
\multicolumn{6}{c }{ Dev-2: Fig.~\ref{sp_dev3}(d) } \\
\hline
$\mathbf{\mathcal{H}_t/\mathcal{H}_{\cpb} }$ &
$\omega_{01}$ & $\omega_{02}$ & $\omega_{03}$ & $\omega_{04}$ & $\omega_{05}$ \\ 
$3^2/10^2$  &4.8741  &5.2474  &5.6737  &9.1045  &9.9481  \\
$4^2/8^2$  &4.8744  &5.2463  &5.6742  &9.1033  &9.9654 \\
$4^2/10^2$  &4.8744  &5.2463  &5.6742  &9.1033  &9.9654  \\
$5^2/12^2$  &4.8745  &5.2462  &5.6743  &9.0133  &9.9621   
\end{tabular}
\end{ruledtabular}
\end{table}

We numerically diagonalize $H_{sys}$~\eqref{hsys}  
in a truncated Hilbert space 
\begin{equation}
\mathcal{H}=\mathcal{H}_A\otimes \mathcal{H}_B \otimes \mathcal{H}_{a}\otimes \mathcal{H}_{b}.
\end{equation}
Therein, $\mathcal{H}_{A}$ is spanned by the lower energy 
eigenstates of the bare transmon Hamiltonian:
\begin{equation}
H_{A}=2e^2\bm{C}^{-1}_{A,A}\hat{n}_A^2-E_{J,t}\cos(\hat{\varphi}_A)
\end{equation}
and similarly for $\mathcal{H}_B$.
The CPB Hilbert space $\mathcal{H}_{a/b}$ is spanned by the charge number states, i.e., the
eigenstates of $\hat{n}_{a/b}$ with eigenvalue $n=-3,-2,\cdots, 4$  (the ``sweet spot'' satisfying $0.5<n_{g,a/b}<1$ is chosen). 
The dimensions of each space is varied to ensure the convergence of the eigenvalues of
the ground and low-excitation eigenvalues. As shown in Tab.~\ref{tab_trunc}, 
we find that it is sufficient to take
\begin{equation}
\mathrm{dim} \mathcal{H}_{A/B}=4,\;
\mathrm{dim}\mathcal{H}_{a/b}=8.
\end{equation}
That is, the numerical  diagonalization is implemented in  a 1024-dimensional Hilbert space.
After the diagonalization, we neglect most of the eigenstates and 
keep only the lowest four excited states. These states are then substituted into
the general formula for the scattering matrix, to be introduced later, to calculate the
fidelity of the circulator.

\section{Equivalence between the bi-CPB core and the Josephson junction Ring}
\label{equivalence-3CPB}

\begin{figure}[b]
\centering
\includegraphics[width=0.9\textwidth]{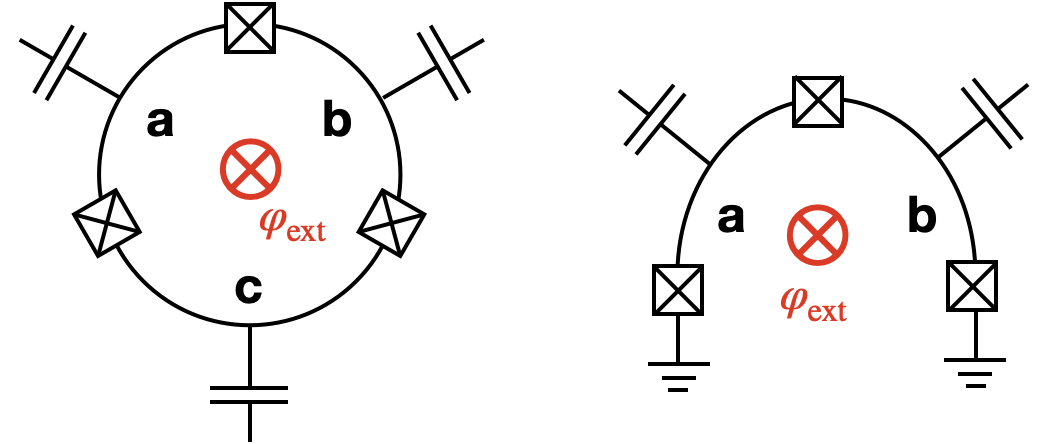}
\caption{Left panel: 3-port Josephson junction ring; Right panel: 2-port bi-CPB core.}
\label{sp_32eq}
\end{figure}

In Fig.~\ref{sp_32eq} we illustrated the 3-port Josephson ring of Ref.~\cite{Koch:2010aa}
and the bi-CPB core used in the main text.
Here we show the equivalence between them.

We derive the Hamiltonian of the Josephson ring following Ref.~\cite{Koch:2010aa}.
The total charge of the three CPBs,
$\hat{N}=\hat{n}_a+\hat{n}_b+\hat{n}_c$, is conserved. Thus, we restrict 
the description to the subspace spanned by states $\ket{\psi}$ satisfying 
$\braket{\psi | \hat{N} |\psi}=N_0$. In this subspace, we apply
the following canonical transformations
\begin{equation}
\begin{aligned}
\hat{\varphi}_a=\hat{\varphi}'_a+\hat{\varphi}'_c,\,
\hat{\varphi}_b=\hat{\varphi}'_c-\hat{\varphi}'_b,\,
\hat{\varphi}_c=\hat{\varphi}'_c, \\
\hat{n}_a=\hat{n}'_a,\, \hat{n}_b=-\hat{n}'_b,\ \hat{n}_c=-\hat{n}'_a+\hat{n}'_b+\hat{n}'_c,
\end{aligned}
\end{equation}
where $\hat{n}'_c=\hat{N}$ is conserved hence the conjugate variable $\hat{\varphi}'_c$ is 
cyclic. Therefore, the Hamiltonian restricted to this subspace has the form
\begin{equation}
\begin{aligned}
H_{(N_0)}=&4E_{C}\bigg[ (\hat{n}'_a-n'_{g,a})^2+ (\hat{n}'_b+n'_{g,b})^2-\hat{n}'_a \hat{n}'_b\bigg] \\
&-E_{J}\cos(\hat{\varphi}'_a-\frac{\varphi_\ext}{3})-E_{J}\cos(\hat{\varphi}'_b
-\frac{\varphi_\ext}{3}) \\
&-E_{J}\cos(\hat{\varphi}'_a+\hat{\varphi}'_B+\frac{\varphi_\ext}{3})
\end{aligned}
\end{equation}
where the junctions are assumed to be identical for convenience, $n'_{g,a}=\frac{1}{2}(n_{g,a}-n_{g,c}+N_0)$ and 
$n'_{g,b}=\frac{1}{2}(n_{g,b}-n_{g,c}+N_0)$.

Next, we perform the
transformations $\hat{\varphi}'_{a/b}\rightarrow \hat{\varphi}'_{a/b}+\frac{1}{3}\varphi_\ext$,
which is then followed by 
\begin{equation}
\hat{\varphi}'_b\rightarrow -\hat{\varphi}'_b,\;
\hat{n}'_b\rightarrow -\hat{n}'_b.
\end{equation}
With this replacement, the Hamiltonian is brought into the form
\begin{equation}
\begin{aligned}
H_{(N_0)}=&4E_{C}\bigg[ (\hat{n}'_a-n'_{g,a})^2+ (\hat{n}'_b-n'_{g,b})^2+\hat{n}'_a \hat{n}'_b\bigg] \\
&-E_{J}\cos(\hat{\varphi}'_a)-E_{J}\cos(\hat{\varphi}'_b) \\
&-E_{J}\cos(\hat{\varphi}'_a-\hat{\varphi}'_b+\varphi_\ext).
\end{aligned}
\end{equation}
This is formally equivalent to the Hamiltonian of our bi-CPB with 
 for suitable charging energies and  offset charges $\tilde{n}_{g,a/b}$ given by
\begin{equation}
\begin{aligned}
\tilde{n}_{g,a} &=\frac{4}{3}n'_{g,a}-\frac{2}{3}n'_{g,b},\\
\tilde{n}_{g,b} &=\frac{4}{3}n'_{g,b}-\frac{2}{3}n'_{g,a}.
\end{aligned}
\end{equation}

\section{Single Photon Scattering Matrix}\label{scattering matrix}
Here we derive the scattering matrix for a single incident photon. 
The formulas derived below will be used to calculate the fidelity of a circulator based on the chiral interfaces with and without the transmons. It will be convenient to expand the operators of the chiral system on
eigenstates of its Hamiltonian, which can then be specified on demand. 
We will apply the rotating-wave approximation to the 
interaction between the chiral system and the transmission line waveguide~\eqref{hint}. This means that an operator representing the emission of a photon is always associated with
a jump from states of higher energy to lower, and vice versa. For simplicity, we also omit the inelastic scattering where a single
input photon is scattered into two or more photons of lower frequency. 
Thus, we consider only the decay from an excited state to the ground state excluding intermediate lower excited state. This is justified in the discussion section of the main text. 
In the treatment below we include the Lamb shift caused by coupling to the waveguide but neglect e.g. effects caused by divergences of the high frequency light modes and corrections to the dipole approximation of the transmon-waveguide coupling.  Such effect should  be absorbed into the model presented below by renormalizing parameters like energy levels and coupling strengths. A more sophisticated theory which carefully addresses all these effects is beyond of the scope of this work.

Explicitly, the system Hamiltonian is expressed as
\begin{widetext}
\begin{equation}\label{seq-SH}
H/\hbar=\int_{-\pi}^{\pi}\frac{dk}{2\pi}\, \omega_k b_k^{\dagger} b_k
+\sum_{n}\, e_n\ket{\phi_n}\bra{\phi_n} 
+i\sum_{\xi,i;n}\int_{-\pi}^{\pi}\frac{dk}{2\pi} \sqrt{\kappa_{\xi i k}}
b_{k}^{\dagger}e^{-ikx_{\xi}}\eta_{i n}\sigma^{-}_{n}+h.c.
\end{equation}
where index $\xi\in\{ l, r\}$, index $n$ runs over the numerically obtained excited states $\ket{\phi_n}$ of
$H_{sys}$  ($n\leq 4$ in our calculation),
index $i\in\{a,b,A,B\}$,
$\sigma^{-}_{n}=\ket{g}\bra{\phi_n}$ and $\eta_{i n}\equiv\braket{g | \hat{n}_i |\phi_n }$
with $\ket{g}$ the ground state of $H_{sys}$.
Decay between the excited states are neglected, as explained above.
The coupling strength is
\begin{equation}
\sqrt{\kappa_{\xi i k}}=2e\bm{C}^{-1}_{\xi,i}\sqrt{\frac{\abs{\sin(k/2)}}{\hbar Z_0}}.
\end{equation}

To describe a situation where a single photon is incident, we consider the single-excitation ansatz
\begin{equation}
\ket{\Psi}=\sum_{n}c_n\ket{\phi_n}+\int_{-\pi}^{\pi}\frac{dk}{2\pi}f_k b_k^{\dagger}\ket{g,\emptyset},
\end{equation}
where $\ket{\emptyset}$ is the vacuum of the waveguide photon.
Then the Schr{\"o}dinger equation leads to
\begin{subequations}
\begin{equation}\label{eq-cn-d}
i\partial_t c_n=c_n e_n -i\int_{-\pi}^{\pi}\frac{dk}{2\pi} f_k\sum_{\xi,i} e^{ikx_\xi}
\sqrt{\kappa_{\xi i k}}\eta_{i n}^{*},
\end{equation}
\begin{equation}
i\partial_t f_k= \omega_k f_k+i\sum_{\xi, i, n} c_n 
e^{-ik x_\xi}\sqrt{\kappa_{\xi i k}}\eta_{i n}.
\end{equation}
\end{subequations}
The second equation is formally solved in two ways
\begin{subequations}
\begin{equation}\label{input-f-d}
f_k(t)= f_k(t_0)e^{-i\omega_k(t-t_0)} 
+\sum_{\xi,i,n} e^{-ikx_\xi}\sqrt{\kappa_{\xi i k}}\eta_{i n}\int_{t_0}^{t}d\tau e^{-i\omega_k(t-\tau)}c_n(\tau),
\end{equation}
and
\begin{equation}
f_k(t)=  f_k(t_f)e^{-i\omega_k(t-t_f)}
+\sum_{\xi,i,n} e^{-ikx_\xi}\sqrt{\kappa_{\xi i k}}\eta_{i n}
\int_{t_f}^{t}d\tau e^{-i\omega_k(t-\tau)}c_n(\tau).
\end{equation}
\end{subequations}
Therein, the input and output fields are defined as $f^{in}_k(t)=f_k(t_0)e^{-i\omega_k(t-t_0)}$
and $f^{out}_k(t)=f_k(t_f)e^{i\omega_k(t_f-t)}$, respectively. 
In Fourier space (assuming $t_f-t_0=\infty$) we then obtain the input-output relation
\begin{equation}\label{output-input-Cw}
 F^{out}_k(\omega)-F^{in}_k(\omega) 
= 2\pi\delta(\omega-\omega_k)
\sum_{\xi,i,n} e^{-ikx_\xi}\sqrt{\kappa_{\xi i k}}\eta_{i n}C_n(\omega).
\end{equation}
We substitute the input field solution Eq.~\eqref{input-f-d} into Eq.~\eqref{eq-cn-d}
and obtain
\begin{equation}
(i\partial_t - e_n)c_n = -i\int_{-\pi}^{\pi}\frac{dk}{2\pi}
\sum_{\xi,i}e^{ikx_\xi}\sqrt{\kappa_{\xi i k}}\eta^{*}_{i n}f_k^{in}(t) 
-i\sum_{\xi \xi' i i' n'}\int_{-\pi}^{\pi} \frac{dk}{2\pi} e^{ik(x_\xi-x_{\xi'})} \sqrt{\kappa_{\xi i k}\kappa_{\xi' i' k}}
\eta_{i n}^{*}\eta_{i' n'} 
\int_{t_0}^{t}d\tau e^{-i\omega_k(t-\tau)}c_{n'}(\tau).
\end{equation}
The integral of $\tau$ is evaluated as (assuming $t-t_0 \rightarrow \infty$)
\begin{equation}
\int_{t_0}^{t}d\tau e^{-i(\omega_k-\omega_0)(t-\tau)}\tilde{c}_{n'}(\tau) e^{-i\omega_0 (t-t_0)}
\approx  c_{n'}(t)\frac{i}{\omega_0-\omega_k+i0^{+}}.
\end{equation}
The fast evolution part of $c_{n'}(t)$ is determined by the input frequency $\omega_0$.
The term $\sqrt{\kappa_{\xi  i k}\kappa_{\xi' i' k}}$ contributes a factor
of $\abs{\sin(k/2)}$. Then the integral over $k>0$ is approximated by defining
$q=k-k_0$ and extending the range of integral over $q$ to infinity
\begin{equation}\label{integral-q}
 e^{ik_0(x_\xi-x_{\xi'})} \abs{\sin\frac{k_0}{2}} 
\int_{-\infty}^{\infty}\frac{dq}{2\pi} e^{iq(x_\xi-x_{\xi'})} 
\bigg(\mathcal{P}\frac{-i}{v_g q}+\pi\delta(v_g q)\bigg).
\end{equation}
For the principal integral, we substitute $e^{iq\Delta x}=\cos(q\Delta x)+i\sin(q\Delta x)$ 
and notice that terms of cosine vanish
and those of sine can be solved by
$\int_{-\infty}^{\infty} dx\frac{\sin x}{x}=\pi$.
That is, the integral over $q$ of Eq.~\eqref{integral-q} vanishes if 
$x_\xi-x_{\xi'}<0$, equals $\frac{1}{v_g}$ if $x_\xi-x_{\xi'}>0$, and
equals $\frac{1}{2v_g}$ if $x_\xi=x_{\xi'}$. As to the integral of $k<0$, it is approximated by, defining 
$k=-k_0+q$,
\begin{equation}
 e^{-ik_0(x_\xi-x_{\xi'})}\abs{\sin\frac{k_0}{2}}
\int_{-\infty}^{\infty}\frac{dq}{2\pi} e^{iq(x_{\xi}-x_{\xi'})} 
\bigg(\mathcal{P}\frac{i}{v_gq}+\pi\delta(v_gq)\bigg).
\end{equation}
Similarly, now it vanishes if $x_{\xi}-x_{\xi'}>0$, equals $\frac{1}{v_g}$ if $x_\xi-x_{\xi'}<0$, 
and equals $\frac{1}{2v_g}$ if $x_{\xi}=x_{\xi'}$.
Then we take it into the integral over $k$ and obtain
\begin{equation}
\int_{-\pi}^{\pi}\frac{dk}{2\pi} e^{ik(x_{\xi}-x_{\xi'})}\frac{i\abs{\sin(k/2)}}{\omega_0-\omega_k+i0^{+}}
\approx  v_g^{-1}\abs{\sin\frac{k_0}{2}}e^{ik_{0}\abs{x_\xi-x_{\xi'}}}
\end{equation}
where $k_{0}=\omega_0/v_g >0$. It leads to
\begin{equation}\label{markov-cn}
(i\partial_t-e_n )c_n = -i\int_{-\pi}^{\pi} \frac{dk}{2\pi} \sum_{\xi,i}e^{ikx_\xi}\sqrt{\kappa_{\xi i k}}
\eta^{*}_{i n}f_k^{in}(t) 
-i\sum_{\xi \xi' i i' n'} \frac{\sqrt{\kappa_{\xi i k_{0}}\kappa_{\xi' i' k_{0}}}}{v_g} 
\eta_{i n}^{*}\eta_{i' n'} e^{ik_{0}\, \abs{x_\xi -x_{\xi'}}} c_{n'}.
\end{equation}
In the Fourier space, it is expressed as
\begin{equation}
C_n(\omega)=-i\int_{-\pi}^{\pi} \frac{dk}{2\pi} 
\bigg[\omega\mathbb{I}-\hat{E}+i\Sigma\bigg]^{-1}_{nm} \bm{\alpha}_m(k) F_k^{in}(\omega)
\end{equation}
where $\hat{E}_{nm}=e_n \delta_{n,m}$ and 
\begin{subequations}
\begin{equation}
\bm{\alpha}_m(k)=\sum_{\xi,i}e^{ik x_\xi}\sqrt{\kappa_{\xi i k}}\eta_{i m}^{*}
\end{equation}
\begin{equation}\label{self-energy}
\Sigma_{n,m}=\frac{1}{v_g}\sum_{\xi \xi' i i'}\sqrt{\kappa_{\xi i k_0}\kappa_{\xi' i' k_0 }} 
e^{ik_{0}\abs{x_\xi -x_{\xi'}}} \eta_{i n}^{*}\eta_{i' m}.
\end{equation}
\end{subequations}
Substituting it into Eq.~\eqref{output-input-Cw}, we get
\begin{equation}\label{scattering-matrix}
F_{k_0}^{out}-F_{k_0}^{in}=-i\frac{1}{v_g} \sum_{m,n}
\bm{\alpha}_{n}^{*}(k_0)\bigg[\omega_{k_0}\mathbb{I}-\hat{E}+i\Sigma\bigg]^{-1}_{nm}
[\bm{\alpha}_m(k_0)F_{k_0}^{in}+\bm{\alpha}_{m}(-k_0)F_{-k_0}^{in}]
\end{equation}
This leads to the scattering matrix 
\begin{equation}\label{S-Matrix}
\begin{pmatrix} F_{k_0}^{out} \\ F_{-k_0}^{out} \end{pmatrix}
=\begin{pmatrix}
1-\frac{i}{v_g}\bm{\alpha}_{n}^{*}(k_0)[(\omega_{k_0} -e_n)\mathbb{I}+i\Sigma]^{-1}_{nm}\bm{\alpha}_m(k_0) 
&-\frac{i}{v_g}\bm{\alpha}_{n}^{*}(k_0)[(\omega_{k_0} -e_n)\mathbb{I}+i\Sigma]^{-1}_{nm}\bm{\alpha}_m(-k_0) \\
-\frac{i}{v_g}\bm{\alpha}_{n}^{*}(-k_0)[(\omega_{k_0} -e_n)\mathbb{I}+i\Sigma]^{-1}_{nm}\bm{\alpha}_m(k_0) 
& 1-\frac{i}{v_g}\bm{\alpha}_{n}^{*}(-k_0)[(\omega_{k_0} -e_n)\mathbb{I}+i\Sigma]^{-1}_{nm}\bm{\alpha}_m(-k_0)
\end{pmatrix} 
\begin{pmatrix} F_{k_0}^{in} \\ F_{-k_0}^{in} \end{pmatrix}.
\end{equation}
\end{widetext}
In the evaluation of the scattering matrix, the coupling strengths are related
to the physical parameters according to the  relation
\begin{equation}
\frac{\sqrt{\kappa_{\xi i k }\kappa_{\xi' i' k}}}{v_g}= 
(2e c\delta_z)^2\bm{C}^{-1}_{\xi,i}\bm{C}^{-1}_{\xi', i'}\frac{1}{2\hbar}Z_0 \omega_k.
\end{equation}
The fidelity of the circulator can be evaluated by substituting the matrix elements of Eq.~\eqref{S-Matrix} into Eq.~(2) of the main text.

\section{Additional results for level stability and device tunability}\label{dev-3}
In the main text, the numerical results are obtained for rather symmetric
device parameters, that is, the capacitors used in the bi-CPB core
are assumed to have identical capacitance, and $k_0 d$ and $\varphi_\ext$
are both taken to be $0.5\pi$.
To demonstrate that all these precise choices are not essential,
we here present  parallel calculations for another family of device parameters. 

\begin{figure}[!h]
\centering
\includegraphics[width=0.95\textwidth]{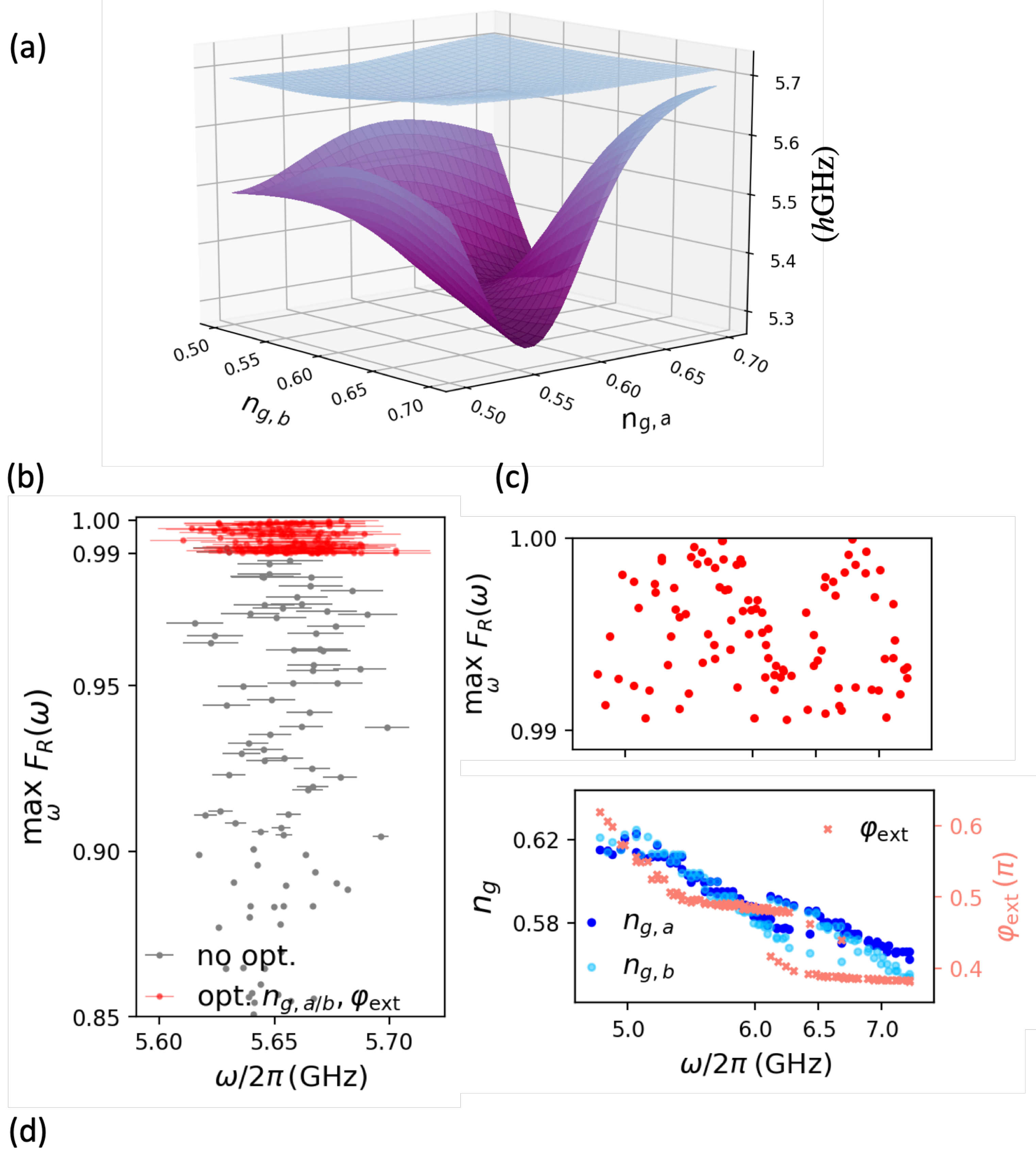}
\caption{
Results parallel to the main text but different device parameters.
(a) The energy levels $\omega_{01}$ and $\omega_{02}$ as a function of
offset charges $n_{g,a(b)}$. (b) 150 samplings of fabrication variations. 
(c) 100 examples of fidelity of chirality obtained by tuning transmon Josephson energies. (d) Device parameters. Units for the capacitances and
the Josephson energies are $\ff$ and $h\ghz$, respectively.
}\label{sp_dev3}
\begin{tabularx}{0.85\textwidth}{c c| c c c c c c | c c c c}
\hline
\hline
$c\delta_z$ &$k_0d$ &$C_q$ &$C_{\Sigma}$ &$C_{ab}$  &$E_{J}$ &$E_{J,ab}$ &$\varphi_{\ext}$
&$C_c$ &$C_t$ &$E_{Jt}$ \\
\hline
50 &0.4$\pi$ &1 &1.5 &0.5 &5.5 &6 &0.49$\pi$  &100  &30 &15\\
\hline
\hline
\end{tabularx}
\end{figure}

Results are plotted in Fig.~\ref{sp_dev3} for  the device parameters
listed in Fig.~\ref{sp_dev3}(d). These plots are equivalent to those
presented in the main text, and they convey the same message.
Our optimization only aims at achieving $\max_\omega F_{R}(\omega)>0.99$. Thus the adapted parameters shown in the lower panel of Fig.~\ref{sp_dev3}(c) are not unique.

\section{Dependence of Fidelity and Bandwidth on the CPB Offset Charges and the External Flux}
\label{more_data}

In the main text, we have shown that fabrication variations can be
addressed by tuning the offset charges $n_{g,a/b}$ and the external flux 
$\varphi_\ext$, but have not gone into details about the optimized values of them.
Since the fabrication variations for capacitance ($\pm 1\%$) is smaller than that of Josephson
energies ($\pm 10\%$), we focus on only the latter
and elaborate on a few examples of extreme variations.

\begin{figure}[t]
\centering
\includegraphics[width=0.95\textwidth]{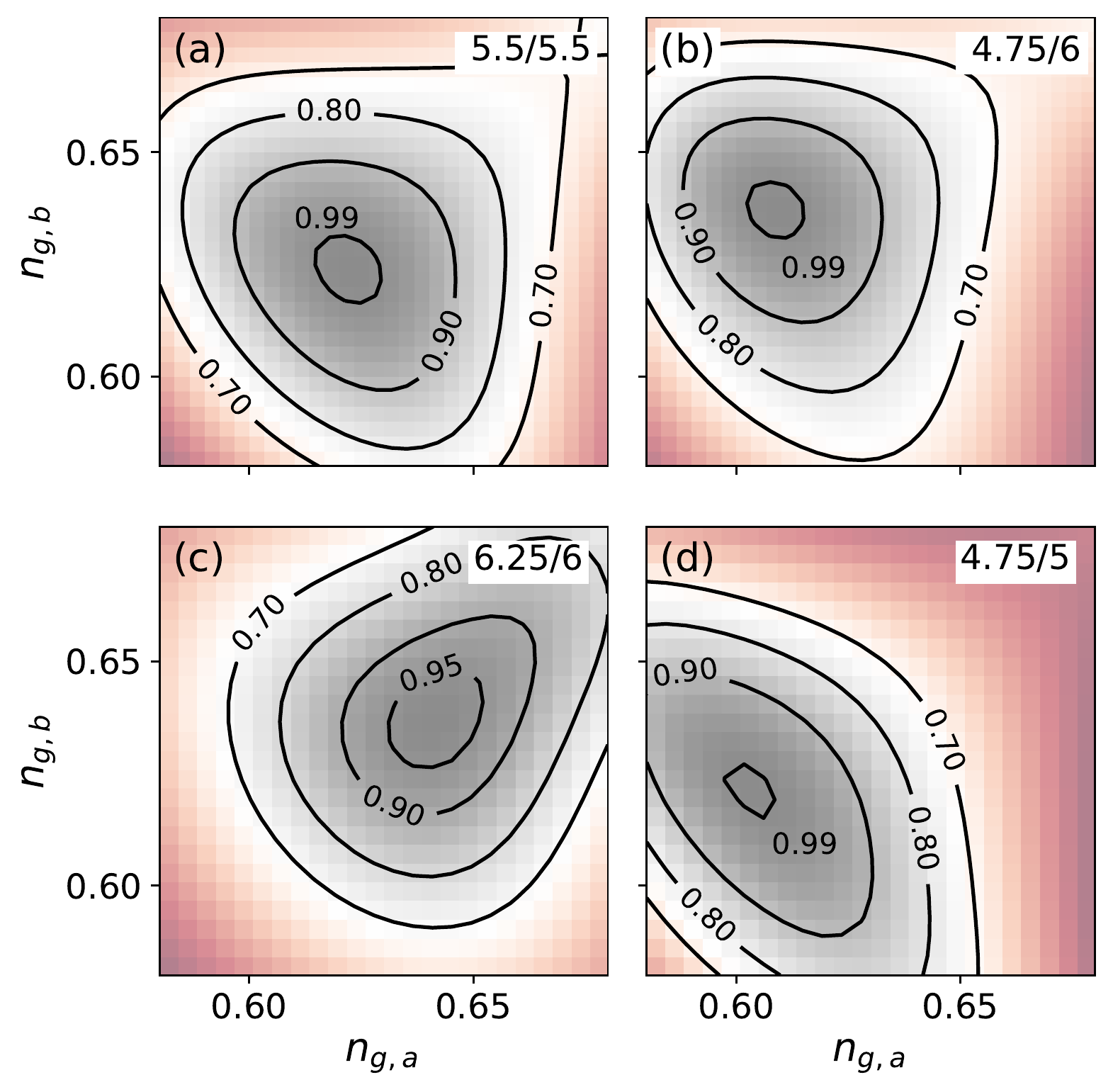}
\caption{ Maximal fidelity as a function of $n_{g,a}$ and $n_{g,b}$ for different values of
$E_{J,a}/E_{J,b}$ labelled in the upper right  corner of each sub-figure (in units of $h\ghz$). The other device
parameters are the same as in Fig.~3(c) of the main text.
}\label{sp_offset}
\end{figure}

We use the device parameters listed in Fig.~3(c) of the main text,
where $E_{J,ab}=5\,h\ghz$ and $E_{J,a}=E_{J,b}=5.5\,h\ghz$. We consider three cases where
$E_{J,a}/E_{J,b}$ deviate from the intended values significantly: one is smaller while the
other is larger, both are larger, or both are smaller, than the intended values.
The peak fidelity of the intended case and three examples of extreme 
fabrication variation are plotted in Fig.~\ref{sp_offset} as a function of
$n_{g,a}$ and $n_{g,b}$, with $\varphi_\ext$ fixed at $0.5\pi$.
It shows that by adjusting only the offset charges,
the fabrication variations of Fig.~\ref{sp_offset}(b) and ~\ref{sp_offset}(d) can be addressed.
In Fig.~\ref{sp_offset}(c) the maximal fidelity cannot be lifted beyond 0.99.
In this case, the external magnetic flux $\varphi_\ext$ must be tuned as well.
We obtained $\max_\omega F_R=0.996$ with parameters
$n_{g,a}=0.63$, $n_{g,b}=0.631$ and $\varphi_\ext=0.534\pi$.

\begin{figure}[!h]
\centering
\includegraphics[width=\textwidth]{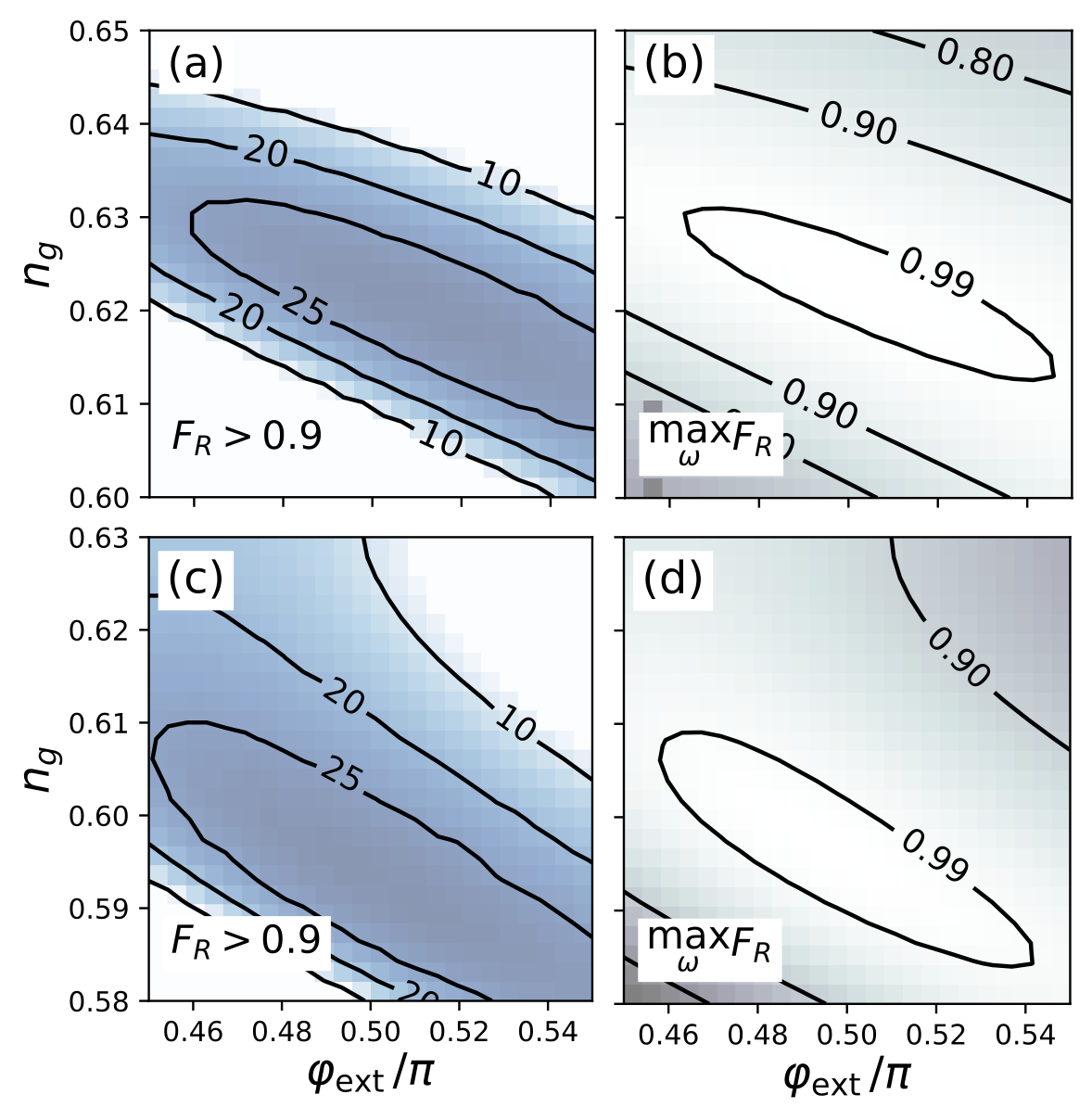}
\caption{ Dependence of the bandwidth for $F_{R}>0.9$ (a,c, in units of $\mathrm{MHz}$) and maximal fidelity $\max_\omega F_{R}(\omega)$ (b,d) on
the offset charge and external flux. The device parameters are: (a,b)
parameter listed in the main text Fig.~3(c);
(c,d) parameter listed in Fig.~\ref{sp_dev3}(d).
}\label{sp_wf}
\end{figure}

Next, we study how the bandwidth and peak fidelity
depend on $n_{g,a/b}$ and $\varphi_\ext$. For convenience,
we assume $n_{g,a}=n_{g,b}=n_g$ and plot the results in Fig.~\ref{sp_wf} for 
the device parameters used in the main text, and those listed in Fig.~\ref{sp_dev3}(d).
From Fig.~\ref{sp_offset} and~\ref{sp_wf} we can see that
to stablize the performance of the chiral interface, the offset charges 
$n_{g, a/b}$ and fluxes should be controlled at the percent level.

\section{Geometric Size and Inherent Loss}\label{size}

Here we estimate the geometric size and inherent loss rate of the proposed photonic chiral interface. For typical devices, the largest component of the system will most likely be the shunt capacitors of the transmons. The lateral length of these capacitors can be as large as $500\mathrm{\mu m}$ in order to
achieve a stronger coupling to the waveguide, such as the devices
reported in Refs.~\cite{Mirhosseini:2019aa} and~\cite{Kannan:2020aa}. (Recall that
in the main text, we use a moderate assumption of $\delta_z=250\mathrm{\mu m}$).
Meanwhile, the length of the transmission line between the two transmons
is about a quarter of the resonant wavelength, which is roughly $1\mathrm{cm}$. The dimension of this can, however, be made much shorter by meandering the wire~\cite{Kannan:2020aa}.
The other components, Cooper-pair-boxes and flux rings of SQUIDs, are typically
down to $10\mathrm{\mu m} \times 10 \mathrm{\mu m}$~\cite{Wallraff:2004aa,Braumuller:2020th}. 
Therefore, the whole device might be placed in a square with
side length being about a few millimeters. This demonstrates the on-chip compatibility of our proposal.

Taking the inherent loss at a rate $\gamma_{in}$ into account,
the highest fidelity of the chirality will be reduced by a small amount 
proportional to $\gamma_{in}/\gamma_{02}$.  Although the Cooper-pair-boxes are sensitive to environmental charge noise, their excitations  are 
strongly suppressed in our ``dark state'' configuration. In our device $\gamma_{in}$ will thus 
mainly be determined by the transmons. Since $T_1$ of a transmon is typically $30\sim 40\mathrm{\mu s}$~\cite{Barends:2013tl}, we expect $\gamma_{in}/\gamma_{02}$ to be well below 0.01, given that $\gamma_{02}\approx 50\sim 100\mathrm{MHz}$.

Next, we give a more precise evaluation of  the  performance of the chiral interface under the influence of  inherent transmon loss at a rate $\gamma_{in}$. Following the approach of Ref.~\cite{Roy:2017tg}, we take the transmon loss into account
by the non-Hermitian Hamiltonian
\begin{widetext}
\begin{equation}
H/\hbar= \int \frac{dk}{2\pi/d}v_g\abs{k} b_k^\dagger b_k+
(\omega_{e}-i\frac{\gamma_{in}}{2})\ket{\phi_e}\bra{\phi_e}
+i\sqrt{\kappa}\int \frac{dk}{2\pi/d} \bigg(
b_k(\sigma^{\dagger}_{A}+e^{ikd}\sigma^{\dagger}_B)-b^\dagger_k(\sigma_{A}+e^{-ikd}\sigma_B)\bigg),
\end{equation}
where $v_g$ is the waveguide group velocity. This non-Hermitian Hamiltonian describe the evolution of the system in the presence of decay and allows us to evaluate the probability for the photon to have the desired behavior through the norm of the wavefunction. 
Compared with Eq.~\eqref{seq-SH} the chiral interface is for simplicity represented by a single  excited state $\ket{\phi_e}=\frac{1}{\sqrt{2}}(\ket{1_A}+e^{i\phi}\ket{1_B})$ with
eigen-energy $\omega_e$. In our chiral photonic interface, 
$\phi$ can be controlled by the off-set charges $n_{g,a/b}$
and the external flux $\phi_{\ext}$, and $\omega_e$
is determined by the tunable transmon frequency.
The transmon-waveguide coupling strength $\sqrt{\kappa}$ is related to the decay rate  to the waveguide through $\Gamma=2\kappa/v_g$.

The same derivations as Sec.~\ref{scattering matrix} leads to the following formula for
the scattering matrix:
\begin{equation}
\begin{pmatrix} F_{k_0}^{out} \\ F_{-k_0}^{out} \end{pmatrix}
=\begin{pmatrix} F_{k_0}^{in} \\ F_{-k_0}^{in} \end{pmatrix}
-\frac{i\Gamma/2}{v_gk_0-\omega_e+i\frac{\gamma_{in}}{2}+i\frac{\Gamma}{2}(1+e^{ik_0d}\cos\phi)}
\begin{pmatrix}
1+\cos(k_0d-\phi) & e^{-ik_0d}(\cos k_0d+\cos\phi) \\
e^{ik_0d}(\cos k_0d+\cos\phi) & 1+\cos(k_0d+\phi)
\end{pmatrix}
\begin{pmatrix} F_{k_0}^{in} \\ F_{-k_0}^{in} \end{pmatrix}.
\end{equation}
\end{widetext}
From above formula we see that there is no reflection for input photons with wavenumber $\pm k_0$
if the phase $\phi$ satisfies $\cos(k_0d)+\cos(\phi)=0$. Equivalently, this leads to the condition
\begin{equation}
\phi=\phi_{L(R)}=\pi+(-)k_0d,
\end{equation}
where $\phi_{L(R)}$ leads to a left(right) chirality. 
Suppose that we consider photon  on resonance with the field corresponding $\omega_e=v_g\abs{k_0}+\frac{1}{4}\Gamma \sin(2k_0 d)$ and $\phi=\phi_R$. Then the
scattering matrix is given by
\begin{equation}
\begin{pmatrix} F_{k_0}^{out} \\ F_{-k_0}^{out} \end{pmatrix}
=\begin{pmatrix}
1-\frac{2\sin^2(k_0 d)}{\epsilon+\sin^2(k_0 d)} & 0 \\
0 & 1
\end{pmatrix}
\begin{pmatrix} F_{k_0}^{in} \\ F_{-k_0}^{in} \end{pmatrix},
\end{equation}
where we have introduced $\epsilon=\gamma_{in}/\Gamma$.
The circulator fidelity to the right $F_R$ is then
\begin{equation}
\begin{aligned}
F_R & =\abs{\frac{\sin^2(k_0d)}{\epsilon+\sin^2(k_0d)}}^2 \\
&\approx 1-2\epsilon\sin^{-2}(k_0 d),
\end{aligned}
\end{equation}
where in the second equality we have assumed that $\epsilon\ll 1$.
The circulator fidelity has the optimal point $1-2\epsilon$ at $k_0d=\pi/2$. 
Recall that the operation frequency of our chiral interface can be tuned in a band of a few GHz.
If the circulator works at $k_0d=0.8\times \pi/2$, the fidelity reads $\approx 1-2.103\epsilon$, which is not much different from the optimum point.

The above calculation shows that the fidelity is limited by the ratio $\epsilon=\gamma_{in}/\Gamma$.
Indeed, superconducting circuit is superior to other platforms of waveguide QED in terms of
lossless $\epsilon$. As an example, in the experiment of Ref.~\cite{Mirhosseini:2019aa}, the transmon Q-1 
 has $\gamma_{in}/2\pi=430\mathrm{kHz}$
and $\Gamma/2\pi=94.1\mathrm{MHz}$, hence $\epsilon=0.0046$ and $\max_{\omega} F_R = 0.991$.
Even better results are feasible for other transmon parameters.  For example, in a recent experiment~\cite{Place:2021ww}, a transmon with lateral size larger than 500 micrometers is fabricated, and found to have a coherence time exceeding $0.3\mathrm{ms}$, i.e., an inherent loss rate of about $\gamma_{in}/2\pi\sim 3\mathrm{kHz}$, yielding a negligible infidelity. 

\bibliography{circuitQED}